\documentclass[fleqn,10pt,twocolumn]{article}
\usepackage[utf8]{inputenc}
\usepackage[T1]{fontenc}
\usepackage{authblk}
\usepackage{cite}   
\usepackage{hyperref}
\usepackage{lineno} 

\usepackage{graphicx}  
\usepackage{subcaption}
\usepackage{multirow}

\usepackage{longtable}
\usepackage[T1]{fontenc}
\usepackage{svg}
\usepackage{bm}        
\usepackage{amsfonts}  
\usepackage{amsmath}   
\usepackage{amssymb}   

\newcommand{\pwisein}{\left\{ \begin{array}{ll}}
\newcommand{\pwiseout}{\end{array}\right.}
\newcommand{\ket}[1]{\left| #1 \right\rangle}

\renewcommand{\det}[1]{\mathrm{det}\left( #1 \right)}

\newcommand{\abs}[1]{\left| #1 \right|}
\newcommand{\ketbra}[2]{\ensuremath{|#1 \vphantom{#2} \rangle \langle #2 \vphantom{#1} | }}

\newcommand{\norm}[1]{\ensuremath{\left| \left|#1 \right| \right|}}
\newcommand{\opmatrix}[3]{\ensuremath{\langle#1 \vphantom{#3} | #2 | #3 \vphantom{#1} \rangle}}
\def\w{\omega}

\author[1,$\dagger$,*]{Andrew R. Cameron}
\author[1,$\dagger$,*]{Jordan M. Thomas}
\author[1]{Alexandru Macridin}
\author[1,2]{Si Xie}
\author[2,3]{Raju Valivarthi}
\author[1]{Soumya S. Ghosh}
\author[5]{Yerko Mu\~{n}oz Barros}
\author[2,3,4]{Neil Sinclair}
\author[1]{Panagiotis Spentzouris}
\author[2,3]{Maria Spiropulu}
\author[6,7]{Prem Kumar}
\author[1]{Cristi\'{a}n Pe\~{n}a}

\affil[1]{Fermi National Accelerator Laboratory, Batavia, IL 60510, USA}
\affil[2]{Division of Physics, Mathematics and Astronomy, California Institute of Technology, Pasadena, CA 91125, USA}
\affil[3]{Alliance for Quantum Technologies (AQT), California Institute of Technology, Pasadena, CA 91125, USA}
\affil[4]{John A. Paulson School of Engineering and Applied Sciences, Harvard University, Cambridge, MA 02138, USA}
\affil[5]{Department of Physics, Universidad Técnica Federico Santa María, Avenida España 1680, 2390123 Valparaíso, Chile}
\affil[6]{Center for Photonic Communication and Computing, ECE Department, Northwestern University, 2145 Sheridan Road, Evanston, Illinois 60208, USA}
\affil[7]{Department of Physics and Astronomy, Northwestern University, 2145 Sheridan Road, Evanston, Illinois 60208, USA}

\affil[$\dagger$]{These authors contributed equally to this work.}
\affil[*]{e-mail: acameron@fnal.gov, jmthomas@fnal.gov}

\date{} 

\begin{document}

\title{Entanglement swapping across a five-node relay in a multiplexed quantum-classical network}
\onecolumn
\begin{abstract} 

Quantum networks are resources for scaling quantum computers and distributed sensing technologies while offering post-quantum security benefits. Teleporting non-classical resources like entanglement, via so called entanglement swapping, is essential for networks in particular overcoming rate-loss limits via quantum repeaters. Deploying these systems on real infrastructure will likely require multiplexing photonic qubits into fibers carrying “classical” light encoding standard Internet communications and control plane signals for multi-node quantum protocols. Here, we report the first demonstration of entanglement swapping and conventional communications operating over the same fibers. Entanglement is swapped across a five-node quantum relay topology connected by four long-distance fibers, each populated with classical data signals. Time-bin entangled photons in the C-band are multiplexed alongside C-band classical signals using dense-wavelength division multiplexing, introducing noise photons generated by high-power classical light. We experimentally and theoretically characterize the trade-off between quantum fidelity and Raman noise photons. Entanglement swapping is demonstrated over a maximum fiber length of 40~km (four 10-km fibers) while simultaneously transmitting 10-Gbps classical data through all fibers. These results represent a significant advancement in the demonstrated complexity of coexisting quantum and classical networks and provide a roadmap for achieving the widespread deployment of advanced quantum technologies.

\end{abstract}
\maketitle

\twocolumn
\section*{Main}

The widespread deployment of quantum optical technology could enable quantum-enhanced encryption, sensing, and distributed quantum computation \cite{kimble_quantum_2008, wehner_quantum_2018}. Building towards large-scale optical fiber-based quantum networks requires accounting for nonideal operating conditions. The existing optical fiber infrastructure is, and will continue to be, dominated by high-power classical traffic, which implies that scalable quantum networking will likely require quantum and classical signals to coexist within the same fiber links. Even with dedicated fiber, quantum networks need to orchestrate quantum distribution, routing, and detection, where classical technologies are often used to handle synchronization, stabilization, and general network management. A central challenge is spontaneous Raman scattering (SpRS) in lit fiber, where strong classical light generates noise photons within quantum passbands with a rate proportional to classical power, wavelength assignment, and fiber length~\cite{chapuran_optical_2009, eraerds_quantum_2010_2, Burenkov:23_2, Thomas:23, thomas_filt}.  

As quantum technology has advanced to increasingly complex protocols, research on ``quantum-classical coexistence'' in fiber networks has necessarily developed alongside~\cite{kumar2026building}, progressing beyond direct transmission of single quantum states~\cite{townsend_simultaneous_1997, chapuran_optical_2009, eraerds_quantum_2010_2, dynes_ultra-high_2016_2, Mao:182, Tbps_CC_narrowfilt_2024} to entanglement distribution~\cite{CC_kumar_ent_1st, Thomas:23, CCband_ent2, 100km, Gul2025_2, OC_entdist_NYC_2025, Talcott2026EntanglementFiber, Wu2026QM_coex}, measurement-device-independent quantum key distribution~\cite{valivarthi_measurement-device-independent_2019_2, berrevoets_deployed_2022}, and quantum state teleportation~\cite{thomasx, song2026quantum}. 

However, entanglement swapping~\cite{zukowski} has only been demonstrated to date in quantum-dedicated fibers~\cite{exp_swap_relay2005, sun_entanglement_2017-1, sun_entanglement_2017-2, Sun2019Nonbilocality, davis2025entanglementswappingsystemsquantum, Craddock2026EntanglementSwapping}. The physics of entanglement swapping~\cite{zukowski}, which enables measurement-induced entanglement between two originally independent particles that never directly interact, underlies many systems ranging from quantum repeater architectures~\cite{repeaters1, repeaters_purification_1999, repeaters, modern_review_repeaters22023} and relays~\cite{relay} to distributed quantum computing~\cite{wehner_quantum_2018, distQC_2014} and network-assisted sensing~\cite{GJC_2012}. Notably, all prior quantum-classical coexistence studies have been limited to at most three connected nodes, whereas many of the envisioned quantum networks above can extend well beyond this.

\begin{figure*}[t]
  \centering
\includegraphics[width=0.9\linewidth]{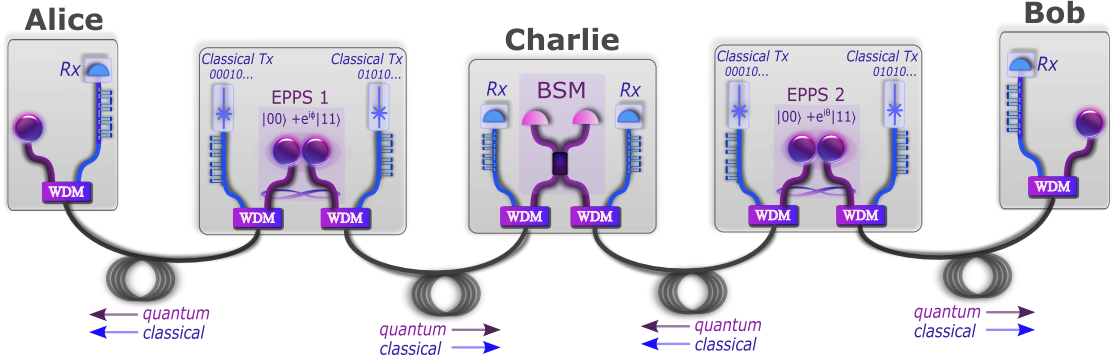}
\setlength{\belowcaptionskip}{-8.5pt} \caption{Conceptual diagram of entanglement swapping across a five-node quantum relay over four fiber links wherein higher-power classical communication signals are wavelength-multiplexed in and out of each fiber. Two independent entangled photon-pair sources (EPPS) generate entangled photon pairs. One photon from each source is transmitted through a fiber segment of length $L/4$ to a central node for a Bell state measurement (BSM). The other photons from the EPPSs are sent a distance $L/4$ to either Alice or Bob.
A successful BSM at Charlie's midpoint relay node heralds entanglement between the two remaining photons at Alice's and Bob's nodes, despite having never directly interacted.}
\label{exp_cartoon}
\end{figure*}

Here, we report the first experimental demonstration of the coexistence of entanglement swapping and classical optical communications over standard long-distance fibers. We demonstrate this in a five-node quantum relay topology with four equal-length fiber segments, which is illustrated in Fig.~\ref{exp_cartoon}. Quantum relays, which have been proposed to increase detector noise tolerance by placing intermediate measurement nodes along a longer-distance fiber~\cite{relay}, also closely resemble a quantum repeater architecture absent from quantum memories~\cite{repeaters1, repeaters_purification_1999, repeaters, modern_review_repeaters22023}.

 We investigate an important scenario where quantum and classical signals both populate the C-band~\cite{CC_kumar_ent_1st, chapuran_optical_2009, eraerds_quantum_2010_2, dynes_ultra-high_2016_2, valivarthi_measurement-device-independent_2019_2, CCband_ent2, Tbps_CC_narrowfilt_2024, thomas_filt}. This choice minimizes attenuation over long distances but substantially increases Raman-induced noise relative to O-band quantum networks~\cite{Mao:182, Thomas:23, thomasx, Talcott2026EntanglementFiber}, allowing us to test the limits of swapping in high-noise environments such as in dense wavelength-division multiplexed (DWDM) integration.
 
We evaluate the performance of entanglement swapping for a range of co-propagating 10-Gbps C-band classical data powers for quantum relays of either 20\,km ($4\times5$~km) or 40\,km ($4\times10$~km) of standard fiber. Notably, all four photons are generated near 1536~nm to be compatible with erbium-ion quantum memory transitions \cite{Er2010, ER2023}. We develop a detailed theoretical model that includes realistic imperfections in swapping systems and the physics of SpRS to simulate our experiment and explore alternative quantum-classical network designs. Importantly, these tools can be readily adapted to inform the impact of other noise sources, including general detector dark counts, background light in free-space communications, or quantum frequency conversion~\cite{yu_QM30km_2020_1324nm, QFC2}. Together, these results provide a significant step toward understanding the challenges of deploying quantum technology toward a quantum internet.

\section{Results}

The conceptual framework of our experiment is illustrated in Figure~\ref{exp_cartoon}. We utilize a quantum relay architecture with five independent quantum nodes, which are equally space along a longer-distance fiber of total length $L$. Entanglement is swapped using two independent entangled photon-pair source (EPPS) which each send one photon to a midpoint Bell state measurement (BSM) node (Charlie) placed at the midpoint of a longer-distance fiber link. 
The two remaining photons are distributed over fiber lengths $L/4$ to outer quantum analyzer nodes (Alice and Bob). Crucially,
all four fibers simultaneously carry wavelength-multiplexed high-power classical telecommunications traffic. A BSM detection at Charlie heralds entanglement between the photons at Alice and Bob, establishing entanglement across the full fiber length $L$.

\begin{figure*}[t]
  \centering
\includegraphics[width=0.9\linewidth]{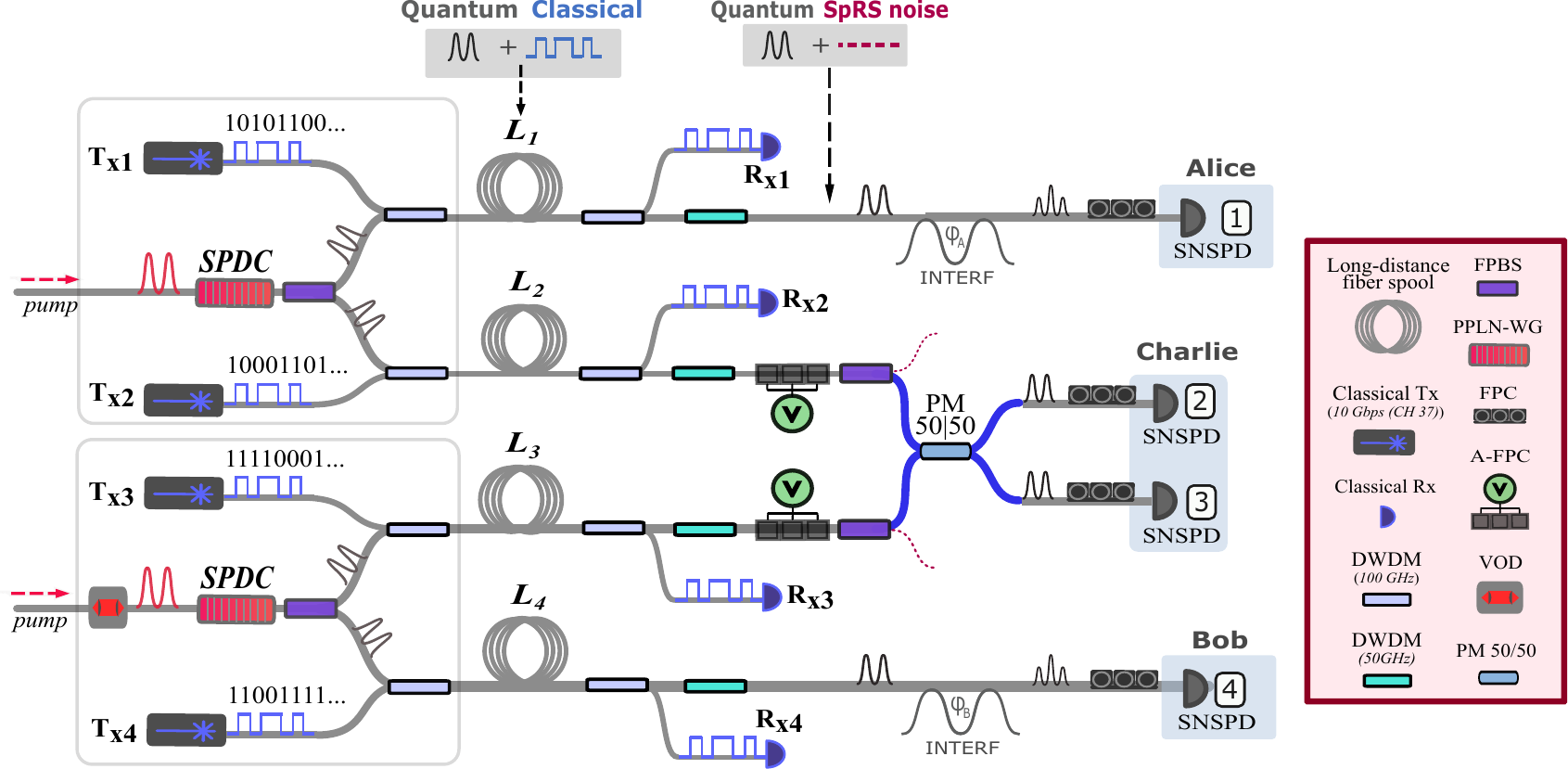}
\setlength{\belowcaptionskip}{-8.5pt} \caption{ Detailed experimental architecture for entanglement swapping in a quantum-classical coexistence relay. The setup utilizes a common 1536.6-nm mode-locked laser to pump two independent SPDC sources. Each of the four quantum paths consists of a SMF-28 fiber spool with either lengths $L=5$~km (20-km total span) or $L=10$~km (40-km total span). Each fiber segment is populated by wavelength-multiplexed C-band time-bin photons co-propagating with 10-Gbps classical data signals allocated to a neighboring DWDM channel. }
\label{exp_setup}
\end{figure*}

The fully fiber-integrated experimental configuration is displayed in Fig.~\ref{exp_setup}. The pump for both entangled photon-pair sources originates from a common 1536.6-nm mode-locked laser with a 6-ps pulse width and a 396.83-MHz repetition rate, which is converted to 768.3-nm pulses using second harmonic generation. This then pumps type-II spontaneous parametric down-conversion (SPDC) to generate wavelength-degenerate photon pairs with a center wavelength of 1536.6~nm, which is designed such that all four photons are near the wavelength of the atomic transitions of erbium-ion quantum memories \cite{Er2010, ER2023}. By using an interferometer to convert the pump into sequential \textit{early} (E) and \textit{late} (L) time bins, SPDC is coherently pumped in each source such that single-photon pair emission encodes the $|\Psi^+\rangle = \frac{1}{\sqrt{2}}(|EE\rangle+|LL\rangle)$ time-bin entangled Bell state (see \nameref{sec:methods}). 
Each of the four 1536.6~nm photons are multiplexed in and out of long-distance SMF-28 fiber spools to co-propagate with 10-Gbps classical light at 1547.7~nm (ITU channel 37) using standard 100-GHz DWDMs.

At Charlie's BSM node, one photon from each source interferes on a polarization-maintaining 50:50 splitter, where each photon is projected onto the $|\Psi^-\rangle = \frac{1}{\sqrt{2}}(|EL\rangle-|LE\rangle)$ Bell state, which swaps the entanglement to the remaining two photons \cite{zukowski}. A high-fidelity BSM requires high single-mode purity as well as indistinguishability in all degrees of freedom. After de-multiplexing the classical light, a narrower 50-GHz channel DWDM with an approximately 41-GHz wide passband is placed in all four quantum paths before single-photon detection. This serves two purposes: to more narrowly filter SpRS as well as to increase the spectral purity of each EPPS for interference at the BSM. 
Over time, polarization and time-of-arrival indistinguishability drift due to temperature and stress changes across the long-distance fibers. To compensate for this, automated fiber polarization controllers (A-FPCs) constantly correct polarization throughout each experiment and time-of-arrival differences between interfering photons are periodically corrected with a variable optical delay (VOD) as outlined in~\nameref{sec:methods}. All photons are measured with superconducting nanowire single-photon detectors (SNSPDs) with $\sim$~90\% detection efficiency and 40-ps full-width half-maximum (FWHM) timing jitter. Coupled with low-jitter time-tagging electronics ($\sim2$~ps), this enables 300-ps coincidence windows to temporally filter SpRS noise that is equally distributed across arrival times~\cite{thomas_filt}. At Alice's and Bob's nodes, Michelson interferometers enable the characterization of time-bin entanglement correlations between them given a successful two-fold coincidence at Charlie's BSM relay node. 

\begin{figure*}[t]
  \centering
\includegraphics[width=\linewidth,trim={1cm 0 6cm 0}, clip]{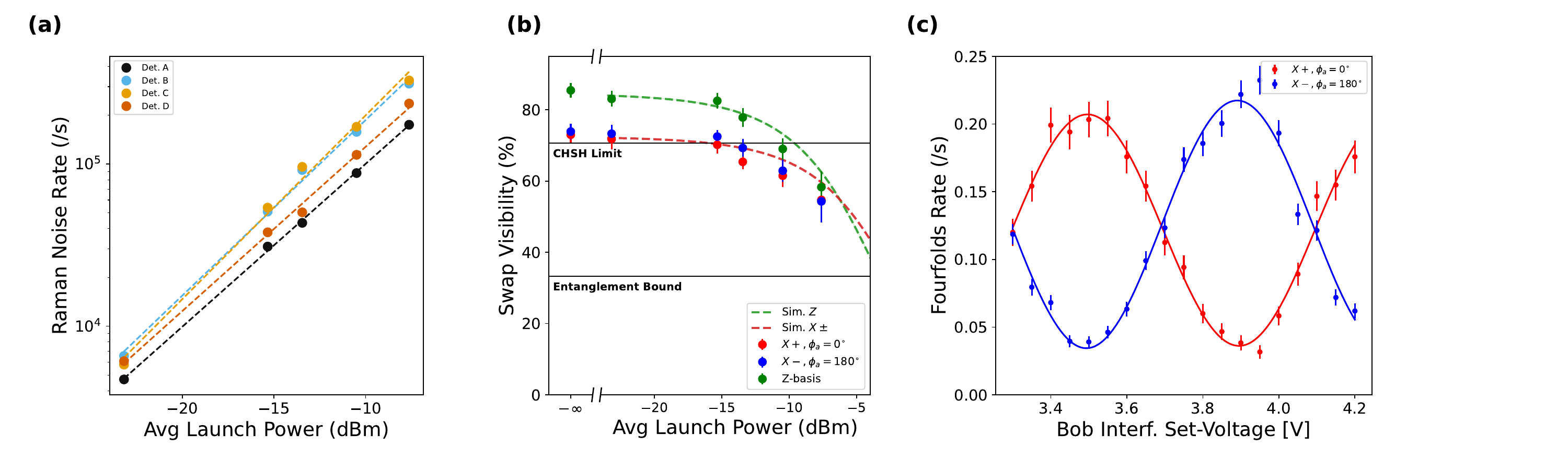}
\setlength{\belowcaptionskip}{-8.5pt} \caption{Experimental results for testing the tolerance of C-band entanglement swapping to SpRS noise caused by co-propagating 10-Gbps
C-band (1547.72~nm) classical data signals in each fiber. In this case, each fiber link is $L/4 = 5$~km, giving a total swapping length of $L= 20$~km. (a) Measured noise counts per second versus the average classical launch power across each fiber link. 
(b) Entanglement swapping visibility versus average classical launch power in each fiber. Experimental data is presented as circular data points and simulation is represented by dashed lines. (c) Four-fold coincidence fringe for the $X^{\pm}$ bases when the received classical power exiting each fiber was $>-18$\,dBm such that both quantum and classical systems were fully operational.}
\label{Results}
\end{figure*}

Entanglement swapping results over a 20-km relay are presented in Fig.~\ref{Results}. Figure~\ref{Results}(a) shows the measured SpRS noise rates in each detector for increasing classical launch powers. The SpRS rates scale linearly with launch power in fiber $j$ as $[\gamma_{1}, \gamma_{2}, \gamma_{3}, \gamma_{4} ] = [1.20, 1.49, 2.61, 1.12]\times 10^6$ counts/s/mW. In the experiment, the fiber lengths and launch powers exhibit a relative variation of $3.6\%$ and $17.5\%$, respectively, due to slight variations in insertion losses and output power. For this reason, we plot our results using the average launch power per fiber ($\langle P_{cl}\rangle$) and vary the power into each fiber equally to ensure consistent trial comparisons. For our specific 10-Gbps transceivers, error-free communication requires $>-18.5$~dBm into the receiver. C-band channels experienced approximately $1.2$-dB attenuation over the 5-km fibers, dictating $P_0>-17.3$~dBm per fiber for full operation.

Figure~\ref{Results}(b) shows the visibility of the swapped entangled state measured in the X and Z bases as we vary $\langle P_{cl}\rangle$. Simulation curves are displayed as dashed lines and are described in detail in Supplementary~\ref{supp_material_numerics}, expanding on previous characteristic function approaches~\cite{davis2025entanglementswappingsystemsquantum, PRXQuantum.1.020317}. Fourfold coincidences are recorded while fixing Alice's measurement interferometer phase ($\phi_a$) while varying Bob's phase ($\phi_b$) for both $\phi_a = 0^{\circ}$ and $180^{\circ}$. In the absence of classical light, we measure swapped entanglement visibilities of $V_X = (73.4 \pm 1.8)\%$ and $V_Z = (85.9 \pm1.4)\%$, as can be seen at the leftmost data point in Fig.~\ref{Results}(b). The slightly reduced X-basis visibility compared to the Z basis arises from imperfect indistinguishability, whereas both are reduced by multiphoton emission \cite{Takeoka_2015}. 

When the C-band classical signals are multiplexed into the fiber links, we observe a monotonic decrease in visibility with increasing launch power due to SpRS noise in each fiber segment. We observe that degradation becomes significant for $\langle P_{cl}\rangle>-15$~dBm. We also observe that the Z basis is more susceptible to noise, which we attribute to the additional 3-dB loss in the measurement interferometers~\cite{thomas_filt}. Figure~\ref{Results}(c) shows an example X-basis coincidence interference fringe between Alice and Bob conditioned on a successful BSM when all classical receivers operated above the minimum error-free threshold. We obtain an average visibility across measurement bases of approximately 75\%, which exceeds the $V>70.7$\% limit required for a Bell test~\cite{CSHS}, and greatly exceeds the $V>33.3$\% entanglement bound~\cite{PhysRevLett.77.1413}.

To characterize the impact of imperfect fidelity excluding SpRS noise, we characterize the quality of each EPPS individually as well as four-fold Hong–Ou–Mandel (HOM) interference visibility~\cite{HOM}, which dictates the quality of the BSM fidelity~\cite{BEll_oper} (see \nameref{sec:methods}). For each source's entanglement visibility without additional background noise, we measure $V_{\rm X, S1} = (95.3 \pm 0.1) \%$ and $V_{\rm X, S2} = (96.6 \pm 0.1)\%$ for sources 1 and 2, respectively. We then characterize four-fold HOM interference between heralded single photons from the independent SPDC sources, achieving a visibility of $V_{\rm HOM} =(85.9 \pm 1.1)\%$. We estimate an indistinguishability of approximately $I=0.92$ between the interfering photons at the BSM when correcting for multiphoton degradation. This shows that the most prominent causes of imperfect fidelity are present in dark fiber. To isolate the degradation due solely to excess noise, we compare the degradation relative to the initial dark fiber visibility. We estimate from Figure~\ref{Results}(b) that an initially ideal swapping setup could sustain $V>70.7$\% up to $\approx-7$~dBm. This demonstrates that the underlying entanglement correlations are highly robust to substantial noise levels.

\begin{figure}[t]
  \centering
\includegraphics[width=\linewidth,trim={0 0 0 0}, clip]{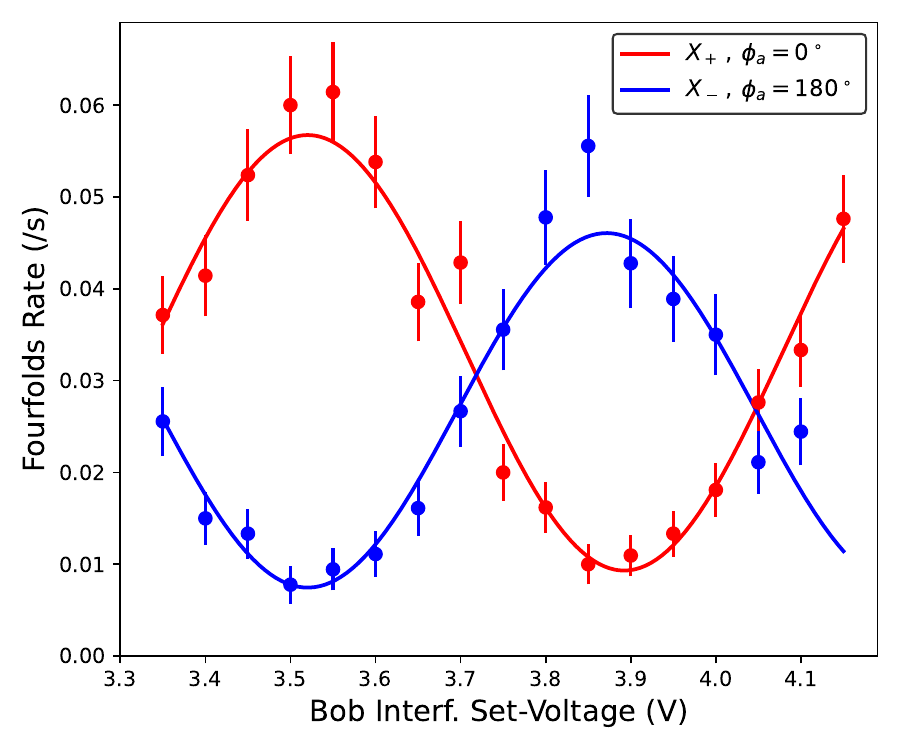}
\setlength{\belowcaptionskip}{-8.5pt} \caption{Visibility of time-bin entanglement swapping across 40\,km of fiber with coexisting 10-Gbps classical signals. Red and blue curves represent separate fringe scans with $\phi_a$ set to $0^{\circ}$ and $180^{\circ}$ by setting the voltage of a static interferometer in Alice's arm.} 
\label{40km_results}
\end{figure}

To demonstrate the scalability of the quantum relay under realistic network conditions, we extended each of the four fiber segments to 10~km, for a total span of 40~km. Doubling the distance adds $\sim1$-dB loss to each link and increases the SpRS interaction length. In dark fiber, we measure swapping visibilities of $V_{X+} = (71.4 \pm 5.1)\%$ and $V_{X-} = (74.1 \pm 4.8)\%$. 

We then turn the classical sources on. Figure~\ref{40km_results} presents the four-fold coincidence rate while the classical sources are transmitted through all four 10-km fibers. The average classical launch power across each spool was set for this measurement to $-18.7$~dBm.
We observed swapped time-bin entangled fringes with visibilities of $V_{X+} =(74.5 \pm 3.8)\%$ and $V_{X-} =(75.0 \pm 4.7)\%$.
In the Z basis, we measured an average visibility of $(85.2 \pm 1.5)\%$. We determine the fidelity to the Bell state $|\Psi^-\rangle$ by the relationship $\mathcal{F} = \frac{1}{4}(1 + 3V_{\rm avg})$~\cite{exp_swap_relay2005}, where we use $V_{\rm avg} = (V_{Z_+} + V_{Z_-} + 2(V_{X+}+V_{X-}))/6$ for the averaged visibility assuming symmetry about the equator of the Bloch sphere. We calculate an average visibility of $(78.2 \pm 2.1)\%$ corresponding to a fidelity of $\mathcal{F}=(83.7 \pm 1.6)\%$. This exceeds the value required for a successful Bell test~\cite{CSHS} by 3.6 standard deviations. This clearly demonstrates the feasibility of simultaneous entanglement swapping and classical communication within real-world long-distance fiber networks. 

\section*{Discussion}

Our results demonstrate that high-fidelity entanglement swapping can be maintained over a minimum of 40~km of optical fiber during simultaneous conventional classical data transmission. Although we demonstrated 10-Gbps classical rates, the power levels achieved in our experiment are sufficient for $>100$-Gbps data rates if replaced by the more power-efficient classical devices used in other demonstrations~\cite{thomasx}.

In Figure~\ref{bidirect}, we use our model (see Supplementary~\ref{supp_material_numerics}) to simulate swapping versus fiber length and alternative quantum-classical wavelength allocation schemes~\cite{thomasSPIE_2}. The results predict that our current system could reach \mbox{$\sim100$~km}, beyond which increasing noise and loss drive a rapid decline in visibility. Alternatively, we plot cases where either quantum or classical signals are in the O-band. The results imply that O-band swapping can achieve the longest distances ($\sim350$~km) because SpRS in the O-band can be nearly five orders of magnitude weaker depending on the wavelength~\cite{Thomas:23, thomasx, Talcott2026EntanglementFiber}, but quantum rates incur an additional 0.13~dB/km loss relative to C-band. Investigating whether O-band quantum repeaters could improve this penalty would be crucial, as inter-city backbone fibers can carry $>20$~dBm of power across the C-band~\cite{Mao:182, Thomas:23, thomasx, Talcott2026EntanglementFiber}. Near-term experimental work will explore O-band swapping, building off the promising high-power coexistence achieved in recent O-band teleportation experiments~\cite{thomasx}. Alternatively, our existing C-band system could tolerate $> 20$~dBm of power in hollow-core fiber~\cite{song2026quantum} due to its ultra-low nonlinearity, but would require new infrastructure to be installed.

Currently, general performance is constrained by source-intrinsic limitations, namely non-unity spectral purity and multiphoton contributions. Addressing these bottlenecks through optimized filter and pump spectral profiles or intrinsically high-purity sources could improve tolerance to classical power by nearly an order of magnitude~\cite{thomas_filt}. These adjustments could enable terabits per second of classical data rates or extend the maximum fiber lengths. 

Early discussions on quantum relays pointed to an inherent distance-scaling advantage for applications limited by detector dark counts~\cite{relay}.
By placing intermediate source and measurement nodes along a longer-distance network route, a higher signal-to-noise ratio at relay nodes can increase the tolerable full link loss relative to direct end-to-end transmission. 
 
\begin{figure}[t!]
\centering
\includegraphics[width=1\linewidth]{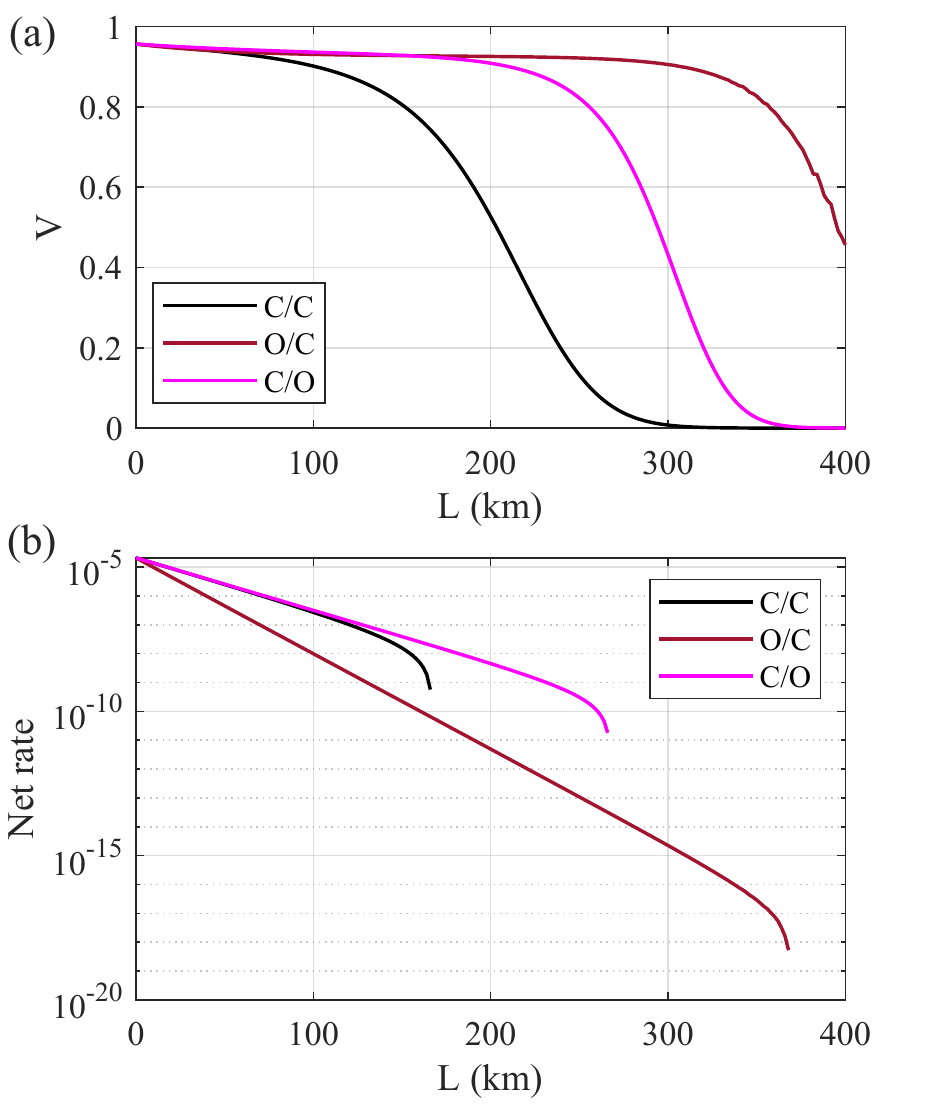}
\caption{Simulation of (a) visibility and (b) coincidence rates versus fiber length for different quantum-classical wavelengths, which modify the SpRS generation and fiber losses. We simulate quantum/classical wavelength combinations of either C-band/C-band (this experiment), C-band/O-band, or O-band/C-band.}
\label{bidirect}
\end{figure}

Interestingly, quantum relays in quantum-classical network environments are predicted to be advantageous over direct communications only in special cases where SpRS is carefully managed~\cite{thomasx}. Since SpRS noise varies with fiber length, intermediate nodes can have notably different levels of noise depending on their location along the full link, whereas dark-fiber relay studies justifiably assume approximately the same intrinsic detector noise-level regardless of location~\cite{relay}. In some configurations, including our four-fiber experimental design, an exact comparison to the performance of direct quantum state transmission is ill-posed. This is because the potential network complexity scales substantially as more photon routes and measurement nodes are included. The performance depends on the number of segments, the length of each fiber, the classical power within them, where each node is located, whether classical signals co- or counter-propagate with the quantum signal, and the wavelength selections for all populated WDM quantum and classical channels. 

For this reason, further exploration of the broad set of scenarios that may arise in complex multi-node quantum-classical networks, including both quantum relays and quantum repeaters, will be important to understand as quantum technology progresses further toward an envisioned quantum internet. This includes quantum memories and entanglement purification that would be needed to extend our system to a full quantum repeater. Rapid experimental progress has been made recently on entangling quantum memories over long-distances~\cite{yu_QM30km_2020_1324nm, QFC2, liu_creation_2024, QM_420km_2026}, but understanding the unique challenges when operating over classical-populated infrastructure has yet to be evaluated. Purification of entanglement is a critical part of quantum repeaters \cite{repeaters_purification_1999, repeaters, modern_review_repeaters22023}. SpRS in long fibers typically results in an approximately basis-independent background noise count for different qubit-encoding types~\cite{Thomas:23, thomasx, thomas_filt}. Some schemes may be more suitable than others for these errors~\cite{purification} and exploring their utility here could further inform the design of high-fidelity quantum networks.

In conclusion, we have demonstrated time-bin entanglement swapping over 40~km of standard fiber in a quantum relay configuration where all four photons propagate through fibers simultaneously carrying classical telecommunications traffic. These results help identify the trade-off between quantum fidelity and classical system power in realistic networks. Future improvements to the classical system and stronger filtering could enable terabits per second of classical data transmission, offering a practical roadmap for deploying scalable quantum-classical networks using telecommunications infrastructure and technologies.

\section*{Methods}
\label{sec:methods}

\subsection*{Photon source and pump preparation}
\label{subsec:pump}

Optical components used to pump each photon source are illustrated in Fig.~\ref{fig:pump_schematic}. A common pulsed laser (Pritel Ultrafast Optical Clock) with center wavelength 1536.6~nm, pulse width of 6~ps, and repetition rate of 396.83~MHz is used to pump both EPPS. A single pulse from the laser is transformed into a double pulse by a Michelson interferometer with a free spectral range of 2.88~GHz to create a pump profile with 346-ps time-bin spacing. A 50:50 splitter breaks the pump into two paths for use in each entangled-pair source. Both sources are amplified by an erbium-doped fiber amplifier (EDFA), then spectrally filtered with a 100-GHz DWDM, and subsequently frequency doubled with a periodically-poled lithium niobate waveguide (PPLN) through second harmonic generation (SHG). After filtering excess unconverted 1536.6-nm light, the resulting 768.3-nm laser pumps a second PPLN for type-II spontaneous parametric down-conversion (SPDC). Generated degenerate photon pairs with a center wavelength of 1536.6~nm (ITU channel 51) have orthogonal polarizations and are separated with a fiber polarizing beam splitter (FPBS). The experiment testing swapping over 20~km operated with mean photon numbers per time bin pulse of $\mu_1 = 0.009$ and $\mu_2 = 0.011$ for Source 1 and Source 2, respectively. To reduce the impact of multiphoton degradation to raw swapping fidelity for the 40-km demonstration, we operated the independent SPDC sources with lower mean photon numbers of $\mu_1 = 0.004$ and $\mu_2 = 0.005$.

The joint spectral intensity (JSI) measurements of each source before spectral filtering are shown in Fig.~\ref{fig:measured_JSIs}. For Source 2, unfiltered and filtered JSIs are compared to illustrate the impact of tight spectral filtering on spectral purity. Tight filtering with 50-GHz bandwidth DWDMs crops the anticorrelated, elongated joint spectrum into a symmetric circular shape, which projects the state closer to a high spectral purity. 

Figure~\ref{fig:source_metrics} shows performance metrics for the two entangled photon sources. Entanglement visibility measurements were conducted by sending each photon to an unbalanced Michelson interferometer with path-length difference matched to the time-bin spacing and post-selecting on the middle bin. One interferometer is held at constant voltage while the other is scanned from $3.25$~V to $4.40$~V. Indistinguishability is determined via HOM interference, where one photon from each source overlaps on a 50:50 beam splitter. All four photons were measured in coincidence by post-selecting on the early bin, and no measurement interferometers were placed in the detection path. A VOD in source 2 as seen in Fig.~\ref{fig:pump_schematic} is scanned to trace out the full dip. 

\subsection*{Classical-quantum multiplexing}
\label{subsec:multiplexing}
Four SMF-28 optical fiber spools contain quantum signals spectrally multiplexed with classical data traffic. All signals are in the C-band and are combined and separated using dense wavelength-division multiplexing (DWDM). The classical signals carry 10-Gbps optical data streams generated and received by small form-factor pluggable (SFP) transceivers operating at 1547.7~nm (ITU channel 37), on the red side of the quantum signal aligned to 1536.6~nm (ITU channel 51). Control and hardware management of the classical transceivers are executed via field-programmable gate arrays programmed using the open source Quantum Instrumentation Control Kit (QICK)~\cite{QICK} platform, running integrated bit error ratio tests (IBERT) for real-time monitoring and measurement of the frame error rate across each active classical channel. Error-free communication for these SFPs required a received power above $-18.5$~dBm.  

To prevent amplified spontaneous emission from the classical transmitters from entering the quantum path, we pre-filter the classical signals before multiplexing into the fiber. We cascade two 100-GHz DWDMs at channel 51 acting as a notch filter at the quantum channel followed by a flat-top passband filter (EXFO) centered around the classical signal's 1547.7-nm center wavelength. For higher input powers, EDFAs were used to amplify the classical signal. Our classical prefiltering also ensured the suppression of EDFA-induced amplified spontaneous emission (ASE). These methods ensure that the measured noise counts in our study are almost entirely due to SpRS generated inside the fibers. 

At the output of each of the four optical fiber spools, demultiplexing is performed using a primary DWDM for the initial separation of quantum and classical signals, and a secondary filtering stage consisting of 50-GHz DWDM filters is applied on each quantum signal path. Both DWDMs provide a combined isolation of out-of-band C-band light to prevent notable noise counts due to leakage from the classical signal's channel. The 50-GHz DWDM serves two additional purposes here. First, the narrower 50-GHz DWDM's passband reduces the total in-band SpRS noise relative to the wider 100-GHz DWDM filters. Second, the filter spectrally carves the joint spectral intensity (JSI) of each photon-pair source to increase the spectral purity of each photon for interference in a BSM. Each 50-GHz channel spacing DWDM has an approximately flat-top passband spectrum that is $\sim$41~GHz wide.

\subsection*{Stabilization and feedback}
\label{subsec:feedback}
The time-of-arrival and polarization of signal photons both drifted throughout the experiments as data was taken on the timescale of hours. Compensation methods for both of these degrees of freedom were run throughout each experiment to maintain high rates in the case of polarization, and high indistinguishability in the case of time-of-arrival. Automated fiber polarization controllers (A-FPCs) placed right before the BSM 50:50 were constantly active and updated every few seconds. Feedback counts were provided to the A-FPCs by Charlie's SNSPDs 2 and 3. The FPBSs right before the BSM 50:50, as seen in Fig.~\ref{exp_setup}, ensure that each interfering photon has the same polarization by filtering out the orthogonal polarization. Therefore, maximizing the singles counts at SNSPDs 2 and 3 by applying the correct voltages across the pins of each A-FPC enabled a real-time polarization correction feedback signal. 

Time-of-arrival corrections were made with the VOD in source 2 and also used single counts at SNSPDs 2 and 3 for feedback. The VOD is placed in the pump path of one source to minimize insertion losses to the quantum paths. To use the timing information from photon counts for feedback into the VOD, we would occasionally interrupt the swapping protocol by delaying one source by 173~ps relative to the other, exactly half of the time-bin spacing. The histograms of the time-stamps relative to our system clock for SNSPDs 2 and 3 are fit to a Gaussian to enable calculation of the arrival time difference of the two photons interfering at the BSM. This process is repeated every 10 minutes and interrupts the protocol for 10 seconds to give time for photon counting and mechanical time scales of the VOD translation. Each calibration cycle achieves a delay-alignment precision of $\sim$ 500 fs RMS. While we note that 10 minutes seemed to be a suitable correction frequency, it is possible that a quickly changing temperature could shift the relative time-of-arrival by a few picoseconds. To minimize the frequency of applying the arrival-time calibration, fiber spools were stored in thermally isolated boxes that maintained a slower drift. With a pump pulse width of $6$~ps, we see high arrival-time indistinguishability at the cost of having very little room for error on time-of-arrival. Higher frequency correction could be applied but would functionally lower rates as the procedure interrupts the protocol. Alternatively, additional SPDs could be added to monitor a tap of the arriving photons for active feedback.

\section*{Acknowledgments}

This work is supported by the Fermi Forward Discovery Group LLC under Contract No. FWP-23-24 with the U.S. Department of Energy, Office of Science, Advanced Scientific Computing Research (ASCR), and the Scientific Enablers of Scalable Quantum Communications program grant (AQNET on Advanced Quantum Networking).

\bibliographystyle{unsrt}
\bibliography{ref}

@article{eraerds_quantum_2010_2,
	title = {Quantum key distribution and 1 {Gbps} data encryption over a single fibre},
	volume = {12},
	issn = {1367-2630},
	doi = {10.1088/1367-2630/12/6/063027},
	 
	number = {6},
	journal = {New Journal of Physics},
	author = {Eraerds, P and Walenta, N and Legré, M and Gisin, N and Zbinden, H},
	month = jun,
	year = {2010},
	pages = {063027},
URL = {https://iopscience.iop.org/article/10.1088/1367-2630/12/6/063027}
}

@article{kimble_quantum_2008,
	title = {The quantum internet},
	volume = {453},
	issn = {0028-0836, 1476-4687},
	url = {http://www.nature.com/articles/nature07127},
	doi = {10.1038/nature07127},
	 
	number = {7198},
	urldate = {2023-04-07},
	journal = {Nature},
	author = {Kimble, H. J.},
	month = jun,
	year = {2008},
	pages = {1023--1030},
}

@article{wehner_quantum_2018,
	title = {Quantum internet: {A} vision for the road ahead},
	volume = {362},
	issn = {0036-8075, 1095-9203},
	shorttitle = {Quantum internet},
	doi = {10.1126/science.aam9288},
	 
	number = {6412},
	journal = {Science},
	author = {Wehner, Stephanie and Elkouss, David and Hanson, Ronald},
	month = oct,
	year = {2018},
	pages = {eaam9288},
  url = {https://www.science.org/doi/10.1126/science.aam9288}
}

@article{townsend_simultaneous_1997,
	title = {Simultaneous quantum cryptographic key distribution and conventional data transmission over installed fibre using wavelength-division multiplexing},
	volume = {33},
	copyright = {© IEE},
	issn = {0013-5194},
	 
	number = {3},
	journal = {Electronics Letters},
	author = {Townsend, P. D.},
	month = jan,
	year = {1997},
	note = {Publisher: Institution of Engineering and Technology},
	pages = {188--190(2)},
URL = {https://digital-library.theiet.org/doi/abs/10.1049/el%3A19970147}
}

@article{chapuran_optical_2009,
	title = {Optical networking for quantum key distribution and quantum communications},
	volume = {11},
	issn = {1367-2630},
	url = {https://iopscience.iop.org/article/10.1088/1367-2630/11/10/105001},
	doi = {10.1088/1367-2630/11/10/105001},
	 
	number = {10},
	urldate = {2025-02-01},
	journal = {New Journal of Physics},
	author = {Chapuran, T E and Toliver, P and Peters, N A and Jackel, J and Goodman, M S and Runser, R J and McNown, S R and Dallmann, N and Hughes, R J and McCabe, K P and Nordholt, J E and Peterson, C G and Tyagi, K T and Mercer, L and Dardy, H},
	month = oct,
	year = {2009},
	pages = {105001},
}

@article{dynes_ultra-high_2016_2,
	title = {Ultra-high bandwidth quantum secured data transmission},
URL ={https://www.nature.com/articles/srep35149},
	volume = {6},
	issn = {2045-2322},
	doi = {10.1038/srep35149},
	 
	number = {1},
	journal = {Scientific Reports},
	author = {Dynes, James F. and Tam, Winci W-S. and Plews, Alan and Fröhlich, Bernd and Sharpe, Andrew W. and Lucamarini, Marco and Yuan, Zhiliang and Radig, Christian and Straw, Andrew and Edwards, Tim and Shields, Andrew J.},
	month = oct,
	year = {2016},
	pages = {35149},
}

@article{Mao:182,
author = {Yingqiu Mao and Bi-Xiao Wang and Chunxu Zhao and Guangquan Wang and Ruichun Wang and Honghai Wang and Fei Zhou and Jimin Nie and Qing Chen and Yong Zhao and Qiang Zhang and Jun Zhang and Teng-Yun Chen and Jian-Wei Pan},
journal = {Opt. Express},
number = {5},
pages = {6010--6020},
publisher = {Optica Publishing Group},
URL = {https://opg.optica.org/oe/fulltext.cfm?uri=oe-26-5-6010&id=382202},
title = {Integrating quantum key distribution with classical communications in backbone fiber network},
volume = {26},
month = {Mar},
year = {2018},
doi = {10.1364/OE.26.006010},
}

@article{Tbps_CC_narrowfilt_2024,
author = {Tianqi Dou and Rende Liu and Shengkai Liao and Jianjun Tang and Jiangang Tong and Rui Ma and Yingxuan Wan and Ruichun Wang and Jun Wu and Xiaolei Zhang and Zhengjun Pan and Yang Li and Chengliang Zhang and Shibiao Tang},
journal = {Opt. Express},
number = {16},
pages = {28356--28369},
publisher = {Optica Publishing Group},
title = {Coexistence of 11 Tbps (110{\texttimes}100 Gbps) classical optical communication and quantum key distribution based on single-mode fiber},
volume = {32},
month = {Jul},
year = {2024},
url = {https://opg.optica.org/oe/abstract.cfm?URI=oe-32-16-28356},
doi = {10.1364/OE.531364}
}

@INPROCEEDINGS{CC_kumar_ent_1st,
  author={Chuang Liang and Kim Fook Lee and Jun Chen and Kumar, P.},
  booktitle={2006 Optical Fiber Communication Conference and the National Fiber Optic Engineers Conference}, 
  title={Distribution of Fiber-Generated Polarization Entangled Photon-Pairs over 100 km of Standard Fiber in OC-192 WDM Environment}, 
  year={2006},
  volume={},
  number={},
  pages={1-3},
  doi={10.1109/OFC.2006.216068},
URL = {https://opg.optica.org/abstract.cfm?uri=ofc-2006-PDP35},
}

@article{Thomas:23,
author = {Jordan M. Thomas and Gregory S. Kanter and Prem Kumar},
journal = {Opt. Express},
number = {26},
pages = {43035--43047},
publisher = {Optica Publishing Group},
title = {Designing noise-robust quantum networks coexisting in the classical fiber infrastructure},
volume = {31},
month = {Dec},
year = {2023},
url = {https://opg.optica.org/oe/abstract.cfm?URI=oe-31-26-43035},
doi = {10.1364/OE.504625},
}

@article{CCband_ent2,
  title = {Energy-time entanglement coexisting with fiber-optical communication in the telecom $C$ band},
URL = {https://journals.aps.org/pra/abstract/10.1103/PhysRevA.108.L020601},
  author = {Fan, Yun-Ru and Luo, Yue and Zhang, Zi-Chang and Li, Yun-Bo and Liu, Sheng and Wang, Dong and Zhang, De-Chao and Deng, Guang-Wei and Wang, You and Song, Hai-Zhi and Wang, Zhen and You, Li-Xing and Yuan, Chen-Zhi and Guo, Guang-Can and Zhou, Qiang},
  journal = {Phys. Rev. A},
  volume = {108},
  issue = {2},
  pages = {L020601},
  numpages = {5},
  year = {2023},
  month = {Aug},
  publisher = {American Physical Society},
  doi = {10.1103/PhysRevA.108.L020601}
}

@article{100km,
author = {A. Rahmouni and P. S. Kuo and Y. S. Li-Baboud and I. A. Burenkov and Y. Shi and M. V. Jabir and N. Lal and D. Reddy and M. Merzouki and L. Ma and A. Battou and S. V. Polyakov and O. Slattery and T. Gerrits},
journal = {J. Opt. Commun. Netw.},
number = {8},
pages = {781--787},
publisher = {Optica Publishing Group},
title = {100-km entanglement distribution with coexisting quantum and classical signals in a single fiber},
volume = {16},
month = {Aug},
year = {2024},
url = {https://opg.optica.org/jocn/abstract.cfm?URI=jocn-16-8-781},
doi = {10.1364/JOCN.518226},
}

@article{Gul2025_2,
  title        = {Noise impact of classical headers on the quantum payload in quantum wrapper networking},
url = {https://opg.optica.org/opticaq/fulltext.cfm?uri=opticaq-3-3-303&id=573274},
  author       = {G{\"u}l, Gamze and Kanter, Gregory S. and Tan, Shannon G. and On, Mehmet Berkay and Proietti, Roberto and Yoo, S.~J.~Ben and Kumar, Prem},
  journal      = {Optica Quantum},
  volume       = {3},
  number       = {3},
  pages        = {303--311},
  year         = {2025},
  doi          = {10.1364/OpticaQ.564321},
}

@ARTICLE{OC_entdist_NYC_2025,
  author={Sena, Matheus and Flament, Mael and Andrewski, Shane and Caltzidis, Ioannis and Bigagli, Niccolo and Rieser, Thomas and Bello Portmann, Gabriel and Sekelsky, Rourke and Braun, Ralf-Peter and Craddock, Alexander N. and Schulz, Maximilian and Jons, Klaus D. and Ritter, Michaela and Geitz, Marc and Holschke, Oliver and Namazi, Mehdi},
  journal={Journal of Optical Communications and Networking}, 
  title={High-fidelity quantum entanglement distribution in metropolitan fiber networks with co-propagating classical traffic}, 
  year={2025},
  volume={17},
  number={12},
  pages={1072-1081},
  doi={10.1364/JOCN.575396}}

@article{Talcott2026EntanglementFiber,
  author = {Talcott, G. and Hess, A. and d'Avossa, L. and Kohlert, S. and Yeh, F. and Chen, J. and Mambretti, J. and Rambo, T. and Kanter, G. and Thomas, J. and Kumar, P.},
  title = {Quantum entanglement distribution coexisting with high-rate, broadband classical optical communications over a real-world fiber connecting remote, synchronized nodes},
  journal = {Optica Quantum},
  year = {2026},
  volume = {4},
  number = {4},
  pages = {342--352},
  doi = {10.1364/OPTICAQ.592786},
  url = {https://doi.org/10.1364/OPTICAQ.592786}
}

@article{valivarthi_measurement-device-independent_2019_2,
	title = {Measurement-device-independent quantum key distribution coexisting with classical communication},
	volume = {4},
	issn = {2058-9565},
	doi = {10.1088/2058-9565/ab2e62},
	 
	number = {4},
	journal = {Quantum Science and Technology},
	author = {Valivarthi, R and Umesh, P and John, C and Owen, K A and Verma, V B and Nam, S W and Oblak, D and Zhou, Q and Tittel, W},
	month = oct,
	year = {2019},
	pages = {045002},
}

@article{sun_entanglement_2017-1,
	title = {Entanglement swapping with independent sources over an optical-fiber network},
	volume = {95},
	issn = {2469-9926, 2469-9934},
	doi = {10.1103/PhysRevA.95.032306},
	 
	number = {3},
	urldate = {2023-04-10},
	journal = {Physical Review A},
	author = {Sun, Qi-Chao and Mao, Ya-Li and Jiang, Yang-Fan and Zhao, Qi and Chen, Si-Jing and Zhang, Wei and Zhang, Wei-Jun and Jiang, Xiao and Chen, Teng-Yun and You, Li-Xing and Li, Li and Huang, Yi-Dong and Chen, Xian-Feng and Wang, Zhen and Ma, Xiongfeng and Zhang, Qiang and Pan, Jian-Wei},
	month = mar,
	year = {2017},
	pages = {032306},
}

@article{berrevoets_deployed_2022,
	title = {Deployed measurement-device independent quantum key distribution and {Bell}-state measurements coexisting with standard internet data and networking equipment},
	volume = {5},
	issn = {2399-3650},
	doi = {10.1038/s42005-022-00964-6},
	number = {1},
	journal = {Communications Physics},
	author = {Berrevoets, Remon C. and Middelburg, Thomas and Vermeulen, Raymond F. L. and Chiesa, Luca Della and Broggi, Federico and Piciaccia, Stefano and Pluis, Rene and Umesh, Prathwiraj and Marques, Jorge F. and Tittel, Wolfgang and Slater, Joshua A.},
	month = jul,
	year = {2022},
	pages = {186},
}

@article{kumar2026building,
  author    = {Kumar, Prem and Thomas, Jordan M.},
  title     = {Building Quantum Networks on Classical Fiber Infrastructure},
  journal   = {Optics and Photonics News},
  volume    = {37},
  number    = {6},
  pages     = {26--33},
  year      = {2026},
  doi       = {10.1364/OPN.37.6.000026},
  url       = {https://doi.org/10.1364/OPN.37.6.000026}
}

@article{Wu2026QM_coex,
  author        = {Wu, Denton and Han, Mingzhe and Wang, Zehao and Ferrari, Ana Luiza and Zalewski, Mika A. and Xie, Yuanheng and Chen, Tingjun and Linke, Norbert M.},
  title         = {Trapped Ion Quantum Networking and Telecommunications Coexisting on One Fiber},
  journal       = {arXiv preprint arXiv:2609.06387},
  year          = {2026},
  eprint        = {2609.06387},
  archivePrefix = {arXiv},
  primaryClass  = {quant-ph},
  doi           = {10.48550/arXiv.2609.06387},
  url           = {https://arxiv.org/abs/2609.06387}
}

@article{thomasx,
author = {Jordan M. Thomas and Fei I. Yeh and Jim Hao Chen and Joe J. Mambretti and Scott J. Kohlert and Gregory S. Kanter and Prem Kumar},
journal = {Optica},
number = {12},
pages = {1700--1707},
publisher = {Optica Publishing Group},
title = {Quantum teleportation coexisting with classical communications in optical fiber},
volume = {11},
month = {Dec},
year = {2024},
url = {https://opg.optica.org/optica/abstract.cfm?URI=optica-11-12-1700},
doi = {10.1364/OPTICA.540362},
}

@article{PhysRevX.2.041010,
  title = {Coexistence of High-Bit-Rate Quantum Key Distribution and Data on Optical Fiber},
  author = {Patel, K. A. and Dynes, J. F. and Choi, I. and Sharpe, A. W. and Dixon, A. R. and Yuan, Z. L. and Penty, R. V. and Shields, A. J.},
  journal = {Phys. Rev. X},
  volume = {2},
  issue = {4},
  pages = {041010},
  numpages = {8},
  year = {2012},
  month = {Nov},
  publisher = {American Physical Society},
  doi = {10.1103/PhysRevX.2.041010},
  url = {https://link.aps.org/doi/10.1103/PhysRevX.2.041010}
}

@article{song2026quantum,
  author        = {Song, Ri-Yao and Zhao, Ya-Zhou and Fan, Yun-Ru and Ma, Yang-Bin and Wei, Yan-Yu and Shen, Si and Li, Hao and You, Li-Xing and Guo, Kai and Guo, Guang-Can and Zhou, Qiang},
  title         = {Quantum teleportation over a field-deployed hollow-core fibre network},
  journal       = {arXiv preprint arXiv:2607.25352},
  year          = {2026},
  doi           = {10.48550/arXiv.2607.25352},
  url           = {https://arxiv.org/abs/2607.25352}
}

@article{zukowski,
  title = {``Event-ready-detectors'' Bell experiment via entanglement swapping},
  author = {\ifmmode \dot{Z}\else \.{Z}\fi{}ukowski, M. and Zeilinger, A. and Horne, M. A. and Ekert, A. K.},
  journal = {Phys. Rev. Lett.},
  volume = {71},
  issue = {26},
  pages = {4287--4290},
  numpages = {0},
  year = {1993},
  month = {Dec},
  publisher = {American Physical Society},
  doi = {10.1103/PhysRevLett.71.4287},
  url = {https://link.aps.org/doi/10.1103/PhysRevLett.71.4287}
}

@article{GJC_2012,
  title = {Longer-Baseline Telescopes Using Quantum Repeaters},
  author = {Gottesman, Daniel and Jennewein, Thomas and Croke, Sarah},
  journal = {Phys. Rev. Lett.},
  volume = {109},
  issue = {7},
  pages = {070503},
  numpages = {5},
  year = {2012},
  month = {Aug},
  publisher = {American Physical Society},
  doi = {10.1103/PhysRevLett.109.070503},
  url = {https://link.aps.org/doi/10.1103/PhysRevLett.109.070503}
}

@article{repeaters,
  title = {Quantum repeaters based on atomic ensembles and linear optics},
  author = {Sangouard, Nicolas and Simon, Christoph and de Riedmatten, Hugues and Gisin, Nicolas},
  journal = {Rev. Mod. Phys.},
  volume = {83},
  issue = {1},
  pages = {33--80},
  numpages = {0},
  year = {2011},
  month = {Mar},
  publisher = {American Physical Society},
  doi = {10.1103/RevModPhys.83.33},
  url = {https://link.aps.org/doi/10.1103/RevModPhys.83.33}
}

@article{sun_entanglement_2017-2,
	title = {Entanglement swapping over 100 km optical fiber with independent entangled photon-pair sources},
	volume = {4},
	issn = {2334-2536},
	doi = {10.1364/OPTICA.4.001214},
	 
	number = {10},
	journal = {Optica},
	author = {Sun, Qi-Chao and Jiang, Yang-Fan and Mao, Ya-Li and You, Li-Xing and Zhang, Wei and Zhang, Wei-Jun and Jiang, Xiao and Chen, Teng-Yun and Li, Hao and Huang, Yi-Dong and Chen, Xian-Feng and Wang, Zhen and Fan, Jingyun and Zhang, Qiang and Pan, Jian-Wei},
	month = oct,
	year = {2017},
	pages = {1214},
}

@article{Sun2019Nonbilocality,
  author  = {Sun, Qi-Chao and Jiang, Yang-Fan and Bai, Bing and Zhang, Weijun and Li, Hao and Jiang, Xiao and Zhang, Jun and You, Lixing and Chen, Xianfeng and Wang, Zhen and Zhang, Qiang and Fan, Jingyun and Pan, Jian-Wei},
  title   = {Experimental demonstration of non-bilocality with truly independent sources and strict locality constraints},
  journal = {Nature Photonics},
  year    = {2019},
  volume  = {13},
  number  = {10},
  pages   = {687--691},
  doi     = {10.1038/s41566-019-0502-7},
  url     = {https://doi.org/10.1038/s41566-019-0502-7}
}

@article{Craddock2026EntanglementSwapping,
  author = {Alexander N. Craddock and Tyler Cowan and Niccolò Bigagli and Suresh Yekasiri and Dylan Robinson and Gabriel Bello Portmann and Aditya Verma and Ziyu Guo and Michael Kilzer and Jiapeng Zhao and Mael Flament and Javad Shabani and Reza Nejabati and Mehdi Namazi},
  title = {High-rate Scalable Entanglement Swapping Between Remote Entanglement Sources on Deployed New York City Fibers},
  journal = {arXiv preprint arXiv:2602.15653},
  year = {2026},
  eprint = {2602.15653},
  archivePrefix = {arXiv},
  primaryClass = {quant-ph},
  doi = {10.48550/arXiv.2602.15653},
  url = {https://arxiv.org/abs/2602.15653}
}

@article{relay,
author = {Daniel Collins and Nicolas Gisin and Hugues De Riedmatten},
title = {Quantum relays for long distance quantum cryptography},
journal = {Journal of Modern Optics},
volume = {52},
number = {5},
pages = {735-753},
year = {2005},
publisher = {Taylor & Francis},
doi = {10.1080/09500340412331283633},
URL = { 
    
        https://doi.org/10.1080/09500340412331283633
    },
eprint = { 
    
        https://doi.org/10.1080/09500340412331283633
    
    

}

}

@inproceedings{thomas_ofc_2023_2,
author = { Thomas, Jordan M and Kanter, Gregory S and Xie, Si and Chung, Joaquin and Valivarthi, Raju and Peña, Cristián and Kettimuthu, Rajkumar and Spentzouris, Panagiotis and Spiropulu, Maria and Kumar, Prem},
booktitle = {Optical Fiber Communication Conference (OFC) 2023},
journal = {Optical Fiber Communication Conference (OFC) 2023},
pages = {Tu3I.3},
publisher = {Optica Publishing Group},
title = {Optimization of Classical Light Wavelengths Coexisting with C-band Quantum Networks for Minimal Noise Impact},
year = {2023},
url = {https://opg.optica.org/abstract.cfm?uri=OFC-2023-Tu3H.3},
doi = {10.1364/OFC.2023.Tu3I.3}
}

@article{Burenkov:23_2,
author = {Ivan A. Burenkov and Alexandra Semionov and  Hala and Thomas Gerrits and Anouar Rahmouni and DJ Anand and Ya-Shian Li-Baboud and Oliver Slattery and Abdella Battou and Sergey V. Polyakov},
journal = {Opt. Express},
number = {7},
pages = {11431--11446},
publisher = {Optica Publishing Group},
title = {Synchronization and coexistence in quantum networks},
volume = {31},
month = {Mar},
year = {2023},
doi = {10.1364/OE.480486},
url = {https://opg.optica.org/oe/fulltext.cfm?uri=oe-31-7-11431&id=528385},
}

@article{Er2010,
  title = {Telecommunication-Wavelength Solid-State Memory at the Single Photon Level},
  author = {Lauritzen, Bj\"orn and Min\'a\ifmmode \check{r}\else \v{r}\fi{}, Ji\ifmmode \check{r}\else \v{r}\fi{}\'{\i} and de Riedmatten, Hugues and Afzelius, Mikael and Sangouard, Nicolas and Simon, Christoph and Gisin, Nicolas},
  journal = {Phys. Rev. Lett.},
  volume = {104},
  issue = {8},
  pages = {080502},
  numpages = {4},
  year = {2010},
  month = {Feb},
  publisher = {American Physical Society},
  doi = {10.1103/PhysRevLett.104.080502},
  url = {https://link.aps.org/doi/10.1103/PhysRevLett.104.080502}
}

@article{ER2023,
  title={Quantum storage of entangled photons at telecom wavelengths in a crystal},
  author={Jiang, Ming-Hao and Xue, Wenyi and He, Qian and An, Yu-Yang and Zheng, Xiaodong and Xu, Wen-Jie and Xie, Yu-Bo and Lu, Yanqing and Zhu, Shining and Ma, Xiao-Song},
  journal={Nature Communications},
  volume={14},
  pages={6995},
  year={2023},
  doi={10.1038/s41467-023-42741-1},
  url={https://www.nature.com/articles/s41467-023-42741-1}
}

@article{QICK,
    author = {Stefanazzi et al., Leandro},
    title = {The QICK (Quantum Instrumentation Control Kit): Readout and control for qubits and detectors},
    journal = {Review of Scientific Instruments},
    volume = {93},
    number = {4},
    pages = {044709},
    year = {2022},
    month = {04},
    issn = {0034-6748},
    doi = {10.1063/5.0076249},
    url = {https://doi.org/10.1063/5.0076249},
    
}

@article{thomas_filt,
  title = {Optimal filtering and generation of entangled photons for quantum applications in the presence of noise},
  author = {Thomas, Jordan M. and Cameron, Andrew R. and Pathiranage, Akil and Xie, Si and Valivarthi, Raju and Spentzouris, Panagiotis and Spiropulu, Maria and Pe\~na, Cristi\'an and Kumar, Prem},
  journal = {Phys. Rev. Appl.},
  volume = {25},
  issue = {6},
  pages = {064064},
  numpages = {21},
  year = {2026},
  month = {Jun},
  publisher = {American Physical Society},
  doi = {10.1103/qd6n-548v},
  url = {https://link.aps.org/doi/10.1103/qd6n-548v}
}

@inproceedings{thomasSPIE_2,
  title = {Multiphoton interference and quantum teleportation coexisting with classical communications in optical fiber},
url = {https://www.spiedigitallibrary.org/conference-proceedings-of-spie/13391/133910D/Multiphoton-interference-and-quantum-teleportation-coexisting-with-classical-communications-in/10.1117/12.3044032.short},
  author = {Thomas, Jordan M. and Kanter, Gregory S. and Kumar, Prem},
  booktitle = {Quantum Communications and Quantum Imaging XXIII},
  series = {Proceedings of SPIE},
  volume = {13391},
  pages = {13391--28},
  year = {2024},
  publisher = {SPIE},
  doi = {10.1117/12.3001234},
}

@article{HOM,
  title = {Measurement of subpicosecond time intervals between two photons by interference},
  author = {Hong, C. K. and Ou, Z. Y. and Mandel, L.},
  journal = {Phys. Rev. Lett.},
  volume = {59},
  issue = {18},
  pages = {2044--2046},
  numpages = {0},
  year = {1987},
  month = {Nov},
  publisher = {American Physical Society},
  doi = {10.1103/PhysRevLett.59.2044},
  url = {https://link.aps.org/doi/10.1103/PhysRevLett.59.2044}
}

@article{BEll_oper,
  title = {Measurement of the Bell operator and quantum teleportation},
url = {https://journals.aps.org/pra/abstract/10.1103/PhysRevA.51.R1727},
  author = {Braunstein, Samuel L. and Mann, A.},
  journal = {Phys. Rev. A},
  volume = {51},
  issue = {3},
  pages = {R1727--R1730},
  numpages = {0},
  year = {1995},
  month = {Mar},
  publisher = {American Physical Society},
  doi = {10.1103/PhysRevA.51.R1727}
}

@article{CSHS,
  title = {Proposed Experiment to Test Local Hidden-Variable Theories},
  author = {Clauser, John F. and Horne, Michael A. and Shimony, Abner and Holt, Richard A.},
  journal = {Phys. Rev. Lett.},
  volume = {23},
  issue = {15},
  pages = {880--884},
  numpages = {0},
  year = {1969},
  month = {Oct},
  publisher = {American Physical Society},
  doi = {10.1103/PhysRevLett.23.880},
  url = {https://link.aps.org/doi/10.1103/PhysRevLett.23.880}
}

@article{modern_review_repeaters22023,
  title = {Quantum repeaters: From quantum networks to the quantum internet},
  author = {Azuma, Koji and Economou, Sophia E. and Elkouss, David and Hilaire, Paul and Jiang, Liang and Lo, Hoi-Kwong and Tzitrin, Ilan},
  journal = {Rev. Mod. Phys.},
  volume = {95},
  issue = {4},
  pages = {045006},
  numpages = {66},
  year = {2023},
  month = {Dec},
  publisher = {American Physical Society},
  doi = {10.1103/RevModPhys.95.045006},
  url = {https://link.aps.org/doi/10.1103/RevModPhys.95.045006}
}

@article{repeaters_purification_1999,
  title = {Quantum repeaters based on entanglement purification},
  author = {D\"ur, W. and Briegel, H.-J. and Cirac, J. I. and Zoller, P.},
  journal = {Phys. Rev. A},
  volume = {59},
  issue = {1},
  pages = {169--181},
  numpages = {0},
  year = {1999},
  month = {Jan},
  publisher = {American Physical Society},
  doi = {10.1103/PhysRevA.59.169},
  url = {https://link.aps.org/doi/10.1103/PhysRevA.59.169}
}

@article{exp_swap_relay2005,
  title = {Long-distance entanglement swapping with photons from separated sources},
  author = {de Riedmatten, H. and Marcikic, I. and van Houwelingen, J. A. W. and Tittel, W. and Zbinden, H. and Gisin, N.},
  journal = {Phys. Rev. A},
  volume = {71},
  issue = {5},
  pages = {050302(R)},
  numpages = {4},
  year = {2005},
  month = {May},
  publisher = {American Physical Society},
  doi = {10.1103/PhysRevA.71.050302},
  url = {https://link.aps.org/doi/10.1103/PhysRevA.71.050302}
}

@article{yu_QM30km_2020_1324nm,
	title = {Entanglement of two quantum memories via fibres over dozens of kilometres},
	volume = {578},
	issn = {0028-0836, 1476-4687},
	doi = {10.1038/s41586-020-1976-7},
	 
	number = {7794},
	urldate = {2023-04-10},
	journal = {Nature},
	author = {Yu, Yong and Ma, Fei and Luo, Xi-Yu and Jing, Bo and Sun, Peng-Fei and Fang, Ren-Zhou and Yang, Chao-Wei and Liu, Hui and Zheng, Ming-Yang and Xie, Xiu-Ping and Zhang, Wei-Jun and You, Li-Xing and Wang, Zhen and Chen, Teng-Yun and Zhang, Qiang and Bao, Xiao-Hui and Pan, Jian-Wei},
	month = feb,
	year = {2020},
	pages = {240--245},
}

@article{QM_420km_2026,
  title = {Entangling Quantum Memories through a 420 km Long Fiber},
  author = {Luo, Xi-Yu and Wang, Chao-Yang and Zheng, Ming-Yang and Wang, Bin and Liu, Jian-Long and Gao, Bo-Feng and Li, Jun and Yan, Zi and Ke, Qiao-Mu and Teng, Da and Wang, Rui-Chun and Wu, Jun and Huang, Jia and Li, Hao and You, Li-Xing and Xie, Xiu-Ping and Xu, Feihu and Zhang, Qiang and Bao, Xiao-Hui and Pan, Jian-Wei},
  journal = {Phys. Rev. Lett.},
  volume = {137},
  issue = {7},
  pages = {070801},
  numpages = {7},
  year = {2026},
  month = {Aug},
  publisher = {American Physical Society},
  doi = {10.1103/ccd6-rf1s},
  url = {https://link.aps.org/doi/10.1103/ccd6-rf1s}
}

@article{QFC2,
	title = {Long distance multiplexed quantum teleportation from a telecom photon to a solid-state qubit},
	volume = {14},
	issn = {2041-1723},
	url = {https://www.nature.com/articles/s41467-023-37518-5},
	doi = {10.1038/s41467-023-37518-5},
	 
	number = {1},
	urldate = {2024-08-17},
	journal = {Nature Communications},
	author = {Lago-Rivera, Dario and Rakonjac, Jelena V. and Grandi, Samuele and Riedmatten, Hugues De},
	month = apr,
	year = {2023},
	pages = {1889},
}

@article{repeaters1,
  title={Quantum repeaters: the role of imperfect local operations in quantum communication},
  author={Briegel, H-J and D{\"u}r, Wolfgang and Cirac, Juan I and Zoller, Peter},
  journal={Physical Review Letters},
  volume={81},
  number={26},
  pages={5932},
  year={1998},
  publisher={APS}
}

@article{purification,
  title = {Quantum link bootstrapping using a RuleSet-based communication protocol},
  author = {Matsuo, Takaaki and Durand, Cl\'ement and Van Meter, Rodney},
  journal = {Phys. Rev. A},
  volume = {100},
  issue = {5},
  pages = {052320},
  numpages = {13},
  year = {2019},
  month = {Nov},
  publisher = {American Physical Society},
  doi = {10.1103/PhysRevA.100.052320},
  url = {https://link.aps.org/doi/10.1103/PhysRevA.100.052320}
}

@article{Takeoka_2015,
doi = {10.1088/1367-2630/17/4/043030},
url = {https://dx.doi.org/10.1088/1367-2630/17/4/043030},
year = {2015},
month = {apr},
publisher = {IOP Publishing},
volume = {17},
number = {4},
pages = {043030},
author = {Masahiro Takeoka and Rui-Bo Jin and Masahide Sasaki},
title = {Full analysis of multi-photon pair effects in spontaneous parametric down conversion based photonic quantum information processing},
journal = {New Journal of Physics}
}

@article{Branczyk_2010,
doi = {10.1088/1367-2630/12/6/063001},
url = {https://doi.org/10.1088/1367-2630/12/6/063001},
year = {2010},
month = {jun},
publisher = {},
volume = {12},
number = {6},
pages = {063001},
author = {Brańczyk, Agata M and Ralph, T C and Helwig, Wolfram and Silberhorn, Christine},
title = {Optimized generation of heralded Fock states using parametric down-conversion},
journal = {New Journal of Physics}
}

@article{liu_creation_2024,
	title = {Creation of memory–memory entanglement in a metropolitan quantum network},
	volume = {629},
	copyright = {2024 The Author(s), under exclusive licence to Springer Nature Limited},
	issn = {1476-4687},
	url = {https://www.nature.com/articles/s41586-024-07308-0},
	doi = {10.1038/s41586-024-07308-0},
	language = {en},
	number = {8012},
	urldate = {2024-11-19},
	journal = {Nature},
	author = {Liu, Jian-Long and Luo, Xi-Yu and Yu, Yong and Wang, Chao-Yang and Wang, Bin and Hu, Yi and Li, Jun and Zheng, Ming-Yang and Yao, Bo and Yan, Zi and Teng, Da and Jiang, Jin-Wei and Liu, Xiao-Bing and Xie, Xiu-Ping and Zhang, Jun and Mao, Qing-He and Jiang, Xiao and Zhang, Qiang and Bao, Xiao-Hui and Pan, Jian-Wei},
	month = {may},
	year = {2024},
	pages = {579--585},
}

@article{JSAanalyit_Silberhorn,
  title = {Limits on the heralding efficiencies and spectral purities of spectrally filtered single photons from photon-pair sources},
  author = {Meyer-Scott, Evan and Montaut, Nicola and Tiedau, Johannes and Sansoni, Linda and Herrmann, Harald and Bartley, Tim J. and Silberhorn, Christine},
  journal = {Phys. Rev. A},
  volume = {95},
  issue = {6},
  pages = {061803},
  numpages = {8},
  year = {2017},
  month = {Jun},
  publisher = {American Physical Society},
  doi = {10.1103/PhysRevA.95.061803},
  url = {https://link.aps.org/doi/10.1103/PhysRevA.95.061803}
}

@article{freq_ent_schmidt,
  title = {Continuous Frequency Entanglement: Effective Finite Hilbert Space and Entropy Control},
  author = {Law, C. K. and Walmsley, I. A. and Eberly, J. H.},
  journal = {Phys. Rev. Lett.},
  volume = {84},
  issue = {23},
  pages = {5304--5307},
  numpages = {0},
  year = {2000},
  month = {Jun},
  publisher = {American Physical Society},
  doi = {10.1103/PhysRevLett.84.5304},
  url = {https://link.aps.org/doi/10.1103/PhysRevLett.84.5304}
}

@article{RamanModel,
	title = {Quantum secured gigabit optical access networks},
	volume = {5},
	issn = {2045-2322},
	url = {https://www.nature.com/articles/srep18121},
	doi = {10.1038/srep18121},
	language = {en},
	number = {1},
	urldate = {2024-07-26},
	journal = {Scientific Reports},
	author = {Fröhlich, Bernd and Dynes, James F. and Lucamarini, Marco and Sharpe, Andrew W. and Tam, Simon W.-B. and Yuan, Zhiliang and Shields, Andrew J.},
	month = dec,
	year = {2015},
	pages = {18121},
}

@article{distQC_2014,
  title = {Large-scale modular quantum-computer architecture with atomic memory and photonic interconnects},
  author = {Monroe, C. and Raussendorf, R. and Ruthven, A. and Brown, K. R. and Maunz, P. and Duan, L.-M. and Kim, J.},
  journal = {Phys. Rev. A},
  volume = {89},
  issue = {2},
  pages = {022317},
  numpages = {16},
  year = {2014},
  month = {Feb},
  publisher = {American Physical Society},
  doi = {10.1103/PhysRevA.89.022317},
  url = {https://link.aps.org/doi/10.1103/PhysRevA.89.022317}
}

@misc{davis2025entanglementswappingsystemsquantum,
      title={Entanglement swapping systems toward a quantum internet}, 
      author={Samantha I. Davis and Raju Valivarthi and Andrew Cameron and Cristian Pena and Si Xie and Lautaro Narvaez and Nikolai Lauk and Chang Li and Kelsie Taylor and Rahaf Youssef and Christina Wang and Keshav Kapoor and Boris Korzh and Neil Sinclair and Matthew Shaw and Panagiotis Spentzouris and Maria Spiropulu},
      year={2025},
      eprint={2503.18906},
      archivePrefix={arXiv},
      primaryClass={quant-ph},
      howpublished  = {arXiv preprint arXiv:2503.18906},
      url={https://arxiv.org/abs/2503.18906}, 
}

@article{PRXQuantum.1.020317,
  title = {Teleportation Systems Toward a Quantum Internet},
  author = {Valivarthi, Raju and Davis, Samantha I. and Pe\~na, Cristi\'an and Xie, Si and Lauk, Nikolai and Narv\'aez, Lautaro and Allmaras, Jason P. and Beyer, Andrew D. and Gim, Yewon and Hussein, Meraj and Iskander, George and Kim, Hyunseong Linus and Korzh, Boris and Mueller, Andrew and Rominsky, Mandy and Shaw, Matthew and Tang, Dawn and Wollman, Emma E. and Simon, Christoph and Spentzouris, Panagiotis and Oblak, Daniel and Sinclair, Neil and Spiropulu, Maria},
  journal = {PRX Quantum},
  volume = {1},
  issue = {2},
  pages = {020317},
  numpages = {16},
  year = {2020},
  month = {Dec},
  publisher = {American Physical Society},
  doi = {10.1103/PRXQuantum.1.020317},
  url = {https://link.aps.org/doi/10.1103/PRXQuantum.1.020317}
}

@article{PhysRevLett.77.1413,
  title = {Separability Criterion for Density Matrices},
  author = {Peres, Asher},
  journal = {Phys. Rev. Lett.},
  volume = {77},
  issue = {8},
  pages = {1413--1415},
  numpages = {0},
  year = {1996},
  month = {Aug},
  publisher = {American Physical Society},
  doi = {10.1103/PhysRevLett.77.1413},
  url = {https://link.aps.org/doi/10.1103/PhysRevLett.77.1413}
}

\clearpage
\onecolumn
\appendix

\section{Supplemental Material - Photon Source Description and Metrics}
\label{supp_material_source}

\begin{figure*}[http]
\centering
\includegraphics[width=1\linewidth]{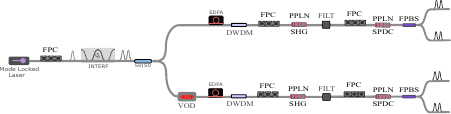}
\caption{Entangled photon source pump preparation schematic. A common mode locked laser pumps two entangled photon-pair sources. A single pulse is converted into a double-pulse shape for time-bin qubit generation via an asymmetric Michelson interferometer (INTERF) with a path length difference of 346~ps. A 50:50 beam splitter splits the pump into two paths, one for each entangled photon source. The two sources are identical except for a variable optical delay (VOD) line for HOM interference timing alignment.}
\label{fig:pump_schematic}
\end{figure*}

\begin{figure*}[http]
\centering
\includegraphics[width=1\linewidth]{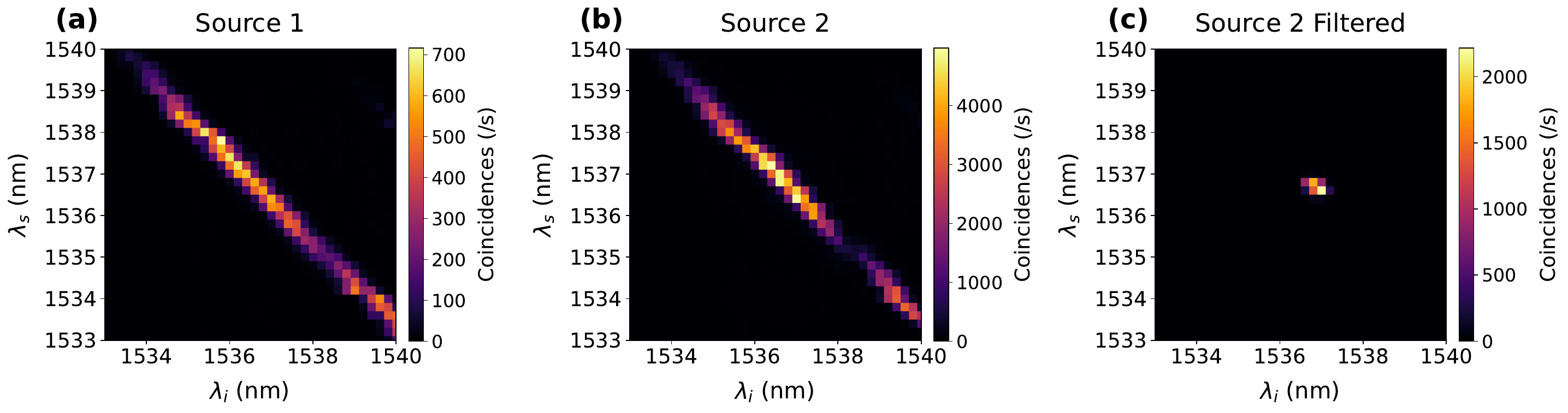}
\caption{Joint spectral intensity (JSI) measurements of time-bin entangled photon pairs generated by spontaneous parametric down-conversion (SPDC). (a,b) Unfiltered degenerate phase-matched signal and idler spectral correlations from Source 1 and Source 2, respectively. (c) Filtered JSI profile from Source 2 obtained by passing signal and idler photons through individual 50-GHz dense wavelength-division multiplexing (DWDM) filters centered at ITU Channel 51 (1536.61~nm). Coincidence rates are reported in coincidences per second (/s).}
\label{fig:measured_JSIs}
\end{figure*}

\begin{figure*}[http]
\centering
\includegraphics[width=1\linewidth]{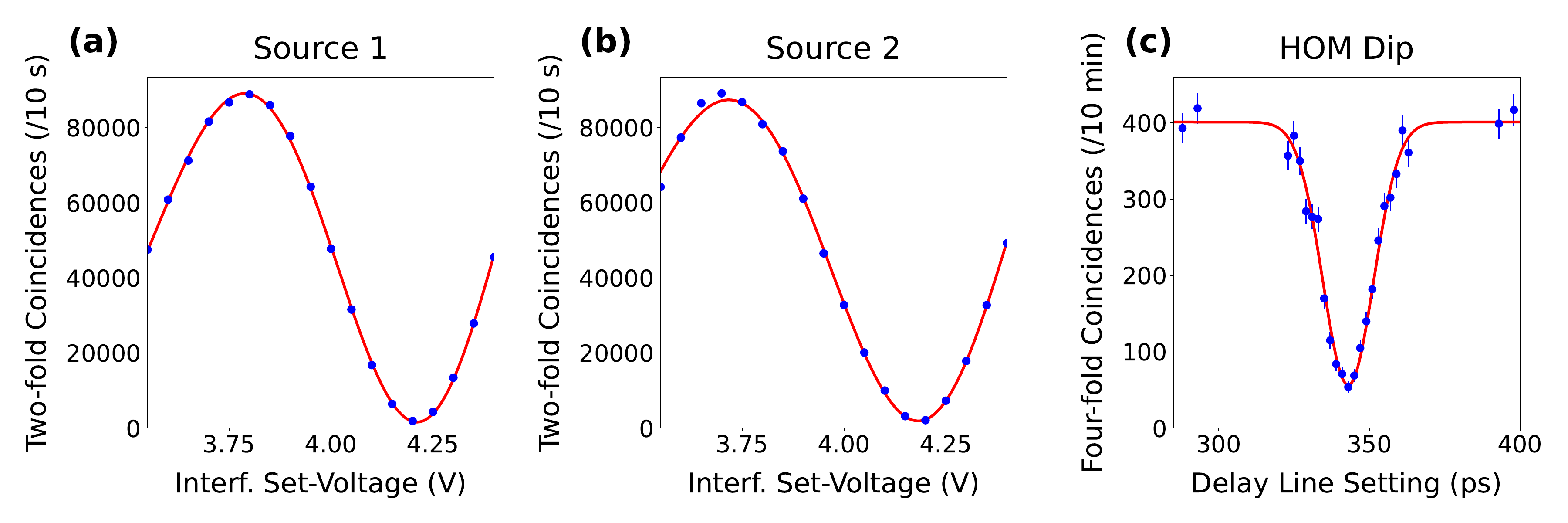}
\caption{(a-b) Entangled photon source characterization. Two-fold coincidence rates as a function of the measurement interferometer set voltage show the $X^+$ two-photon interference fringes of (a) source 1 and (b) source 2, respectively. The visibility is calculated by $V= (V_{\rm max} - V_{\rm max})/(V_{\rm max}-V_{\rm min}).$ Source 1 measured raw visibilities of $V_{Z^+} = (99.20 \pm 0.02) \% $, $V_{Z^-} = (98.17 \pm 0.04) \% $, and $V_{X^+} = (96.57 \pm 0.09) \% $, while source 2 exhibits similar values of $V_{Z^+} = (99.04 \pm 0.02) \% $, $V_{Z^-} = (98.08 \pm 0.04) \% $, and $V_{X^+} = (95.30 \pm 0.10) \% $. All uncertainties are calculated assuming Poisson statistics for photon counting. (c) Hong-Ou-Mandel interference between heralded photons from each source. Fourfold coincidences are recorded while varying an optical delay line to control the degree of indistinguishability. The resulting visibility is $V_{\rm HOM} = (V_{\rm max} - V_{\rm max})/V_{\rm max} =  (86.1 \pm 1.0)\%$. Here, both sources operated with a higher mean photon number near $\mu \approx 0.02$, which contributed a roughly 5\% reduction in visibility. }
\label{fig:source_metrics}
\end{figure*}

\clearpage
\section{Supplemental Material - Numerical Simulations}
\label{supp_material_numerics}

Our numerical simulations model the system evolution downstream of the SPDC sources. Because the state is characterized by a Gaussian Wigner function and all operations preserve Gaussianity, the system can be simulated efficiently within the Gaussian formalism for continuous-variable (CV) quantum systems~\cite{Takeoka_2015}. We extended this formalism to incorporate the spectral and temporal properties of the photons, including the frequency correlations described by the joint spectral amplitude (JSA) of the SPDC sources and a realistic model of optical filtering. Our simulations include a large number of frequency/time modes per arm, ranging from $3,000$ to $10,000$ modes.

For an $N$-mode Gaussian state, the Wigner function is given by
\begin{align}
\label{eq:Wgaussian}
W_{\rho}(q)&= \left(\frac{1}{2 \pi}\right)^N \frac{1}{\sqrt{\det V}}
e^{-\frac{1}{2}\left(q-\bar{q}\right)^T V^{-1} \left(q-\bar{q}\right)},
\end{align}
where $q^T=(x_1,p_1,\dots,x_N,p_N)$, and $x_i$ and $p_i$ are the quadrature coordinates of mode $i$. The state is fully characterized by the $2N\times 2N$ covariance matrix $V$ and the $2N$-dimensional mean vector $\bar{q}$. Gaussian states include the vacuum state, coherent states, and one- and two-mode squeezed states.

Gaussian operations preserve Gaussian states and are completely characterized by their action on the mean vector $\bar{q}$ and the covariance matrix $V$. In particular, transformations generated by Hamiltonians that are at most quadratic in the bosonic mode operators are Gaussian unitaries. Examples of Gaussian operations include common optical operations such as displacements, squeezing, phase rotations, and beam splitters, as well as photon-loss channels and general-dyne measurements.

Mathematically, Gaussian operations are described by the following transformations of the mean vector $\bar{q}$ and covariance matrix $V$:
\begin{align}
\label{eq:gaussiantr1}
\bar{q} &\longrightarrow X\bar{q}+d \\
\label{eq:gaussiantr2}
V &\longrightarrow X V X^T +Y
\end{align}
\noindent where $X$ and $Y$ are real $2N \times 2N$ matrices, and $d$ is a real $2N$-dimensional displacement vector. Unitary Gaussian transformations correspond to the special case $Y=0$, with X being a symplectic matrix. The explicit forms of the matrices X, $Y$, and $d$ corresponding to specific operations can be found in literature, see~\cite{Takeoka_2015} for example. 
In addition to these transformations, partial traces and on–off detection measurements, which model SNSPDs, can be readily incorporated into the formalism. We describe the simulation of these measurements, including the effects of dark noise, in Section~\ref{supp_detectors}.

During the simulation, the Wigner function is represented in three different bases. To accurately describe the correlations encoded in the joint spectral amplitude (JSA), the input state is expressed in the Schmidt basis, which diagonalizes the SPDC interaction Hamiltonian. The filters act as frequency-dependent losses and are implemented in the frequency basis, while the detectors require a representation in the time basis.
Transitions between these bases are described by Gaussian unitary operations.

As can be inferred from Eqs.~\eqref{eq:gaussiantr1}–\eqref{eq:gaussiantr2}, the simulation of Gaussian states reduces to matrix multiplication. The required computational resources scale as $N^3$, and the corresponding operations can be implemented using highly optimized linear algebra libraries. This approach is considerably more efficient than direct simulation in the Fock basis~\cite{Takeoka_2015}, whose computational cost grows exponentially with the number of modes. Moreover, Gaussian-state simulations do not require photon-number truncation, allowing multiphoton effects to be treated exactly within the Gaussian formalism.

\subsection{SPDC output}

The first step of our simulation is to model the joint spectral amplitude $f(\omega_s,\omega_i)$ of the SPDC sources. Experimental measurements of the joint spectral intensity, $|f(\omega_s,\omega_i)|^2$, are used in conjunction with an optimization (or fitting) procedure to determine the parameters $\omega_{s0}, \omega_{i0}, \sigma_p, \sigma_{pm}$, and $\theta$ that characterize the analytical model described by \cite{JSAanalyit_Silberhorn}
\begin{align}
\label{eq:jsamodel}
f(\omega_s,\omega_i)=N \exp\left(\frac{-\left(\omega_s-\omega_{s0}+\omega_i-\omega_{i0}\right)^2}{4\sigma_p^2}\right) \textrm{sinc} \left(\frac{\left(\omega_s-\omega_{s0}\right)\sin\theta+\left(\omega_i-\omega_{i0}\right)\cos\theta}{2\sigma_{pm}}\right).
\end{align}

In the above, $\omega_{j0}$ is the center frequency of the signal ($j=s$) or idler ($j=i$) modes, $\sigma_p$ is the spectral bandwidth of the pump, $\sigma_{pm}$ is the phase-matching bandwidth, and $\theta$ is the phase-matching angle. 

We begin by describing how photon-frequency correlations are incorporated into the simulation. To this end, we use the joint spectral amplitude introduced earlier, which assumes a Gaussian single-pulse laser pumping the SPDC crystal.

We then extend the model to the two-pulse laser configuration required to generate the early and late time bins used in the entanglement-swapping experiment.

\subsubsection{Single-pulse}

The interaction Hamiltonian of an SPDC source is given by
\begin{align}
H_{SPDC}=g\int d\omega_s d\omega_i f(\omega_s,\omega_i) a_s^{\dagger}(\omega_s)a_i^{\dagger}(\omega_i)+H.C.
\end{align}
\noindent where $g$ is the interaction strength, which is proportional to the pump-field amplitude and determines the average number of photons generated by the source. Applying the singular value decomposition (SVD) to the joint spectral amplitude~\cite{freq_ent_schmidt}, 
\begin{align}
f(\omega_s,\omega_i)=\sum_k r_k \xi^s_k(\omega_s)\xi^i_k(\omega_i)
\end{align}
\noindent with $\sum_k r^2_k =1$,
yields the Schmidt modes, which are defined through the creation operators
\begin{align}
\label{eq:schmidtdecomp}
A^{\dagger}_{sk}=\int d\omega \xi^s_k(\omega)a^{\dagger}_s(\omega),~~~~~
A^{\dagger}_{ik}&=\int d\omega \xi^i_k(\omega)a^{\dagger}_i(\omega).
\end{align}

\noindent Since $\xi^s_k(\omega)$ and $\xi^i_k(\omega)$ satisfy the orthogonality relations,
\begin{align}
\int d\omega \xi^{s}_k(\omega)\xi^{s}_j(\omega)=\delta_{kj},~~~~~~\int d\omega\xi^{i}_k(\omega)\xi^{i}_j(\omega)=\delta_{kj},
\end{align}
the Schmidt modes obey the bosonic commutation relations,
\begin{align}
\left[A_{sk}, A_{sk^\prime}^{\dagger}\right]=\delta_{kk^\prime},~~~~~\left[A_{ik}, A_{ik^\prime}^{\dagger}\right]=\delta_{kk^\prime}.
\end{align}

In the Schmidt basis, the SPDC Hamiltonian becomes diagonal,
\begin{align}
H_{SPDC}=g\sum_k r_k A_{sk}^{\dagger} A_{ik}^{\dagger} +H.C.
\end{align}
\noindent and the SPDC output state factorizes into a tensor product of Schmidt two-mode squeezed vacuum states,
\begin{align}
\ket{SPDC}=e^{-iH_{SPDC}t}\ket{0}=\prod_k \otimes \ket{S_k}
\end{align}
\noindent where
\begin{align}
\ket{S_k}=e^{-i g r_k t\left(A_{sk}^{\dagger} A_{ik}^{\dagger}+A_{sk} A_{ik} \right)}\ket{0}.
\end{align}
\noindent The average number of photons in the $k$-th Schmidt mode is given by $\mu_k=\sinh^2 \left(g r_k t\right)$.

Expressing the two-mode squeezed vacuum states in the Fock basis,
\begin{align}
\ket{S_k}=\sqrt{1-\lambda_k^2}\sum_{n=0}^{\infty}
\lambda_k^n \ket{n_{sk},n_{ik}},
\end{align}
\noindent with $\lambda_k = \tanh \left(gr_k t\right)$, the SPDC output can be expanded as
\begin{align}
\ket{SPDC}=\left(\prod_k \sqrt{1-\lambda_k^2}\right)\ket{0}+\sum_k \left(\lambda_k\sqrt{1-\lambda_k^2}\right)A^{\dagger}_{sk}A^{\dagger}_{ik}\ket{0}+\dots {\mathcal{O}}(g^2 t^2)
\end{align}

The single-photon approximation is obtained in the limit of small $g t$ by projecting onto the two-fold coincidence subspace,
\begin{align}
\label{eq:spdc_1photon}
\ket{\phi} \propto \sum_k \left(\lambda_k\sqrt{1-\lambda_k^2}\right)A^{\dagger}_{sk}A^{\dagger}_{ik}\ket{0} \approx g t \sum_k r_k A^{\dagger}_{sk}A^{\dagger}_{ik}\ket{0}
\end{align}

Using the Schmidt decomposition of the joint spectral amplitude Eq~\eqref{eq:schmidtdecomp}, the state in Eq.\eqref{eq:spdc_1photon} can be rewritten in the original frequency basis as
\begin{align}
\ket{\phi} \approx  g t \int d\omega_s d\omega_i f(\omega_s,\omega_i) a_s^{\dagger}(\omega_s)a_i^{\dagger}(\omega_i)\ket{0}.
\end{align}

\subsubsection{Two-pulse pump}

The JSA model described by Eq.\eqref{eq:jsamodel} was derived under the assumption of a single pump pulse with a time-varying Gaussian envelope:
\begin{align}
A_p(t) \propto e^{- t^2  \sigma_p^2}e^{-i \omega_p t}.
\end{align}
The Fourier transform of the pump field yields the Gaussian part of $f(\omega_s,\omega_i)$, {\em{i.e.}} the term
\begin{align}
\exp \left(\frac{-\left(\omega_s-\omega_{s 0}+\omega_i-\omega_{i 0}\right)^2}{4 \sigma_p^2}\right)=\hat{A}_p(\omega_s-\omega_{s0}+\omega_i-\omega_{i0}),
\end{align}
\noindent where $\omega_{s0}+\omega_{i0}=\omega_p$

The formalism can be readily generalized to two (or more) successive pump pulses. For example, two identical pulses separated by a time interval $T$ are described by the pump envelope
\begin{align}
A_{2p}(t)=A_p(t+\frac{T}{2})+ A_p(t-\frac{T}{2}).
\end{align}
Since the Fourier transform of a two-pulse pump is given by
\begin{align}
\hat{A}_{2p}(\omega)=\hat{A}_{p}(\omega)\left(e^{i\frac{\omega T}{2}}+e^{-i\frac{\omega T}{2}}\right)=2 \hat{A}_{p}(\omega) \cos \frac{\omega T}{2},
\end{align}
the two-pulse JSA function is obtained by multiplying the single-pulse JSA by a cosine factor:
\begin{align}
f_{2p}(\omega_s,\omega_i) =2 \cos \frac{\left(\omega_s-\omega_{s0}+\omega_i-\omega_{i0}\right) T}{2} f(\omega_s,\omega_i)
\end{align}

This modification of the JSA function introduces the expected {\em{early}} and {\em{late}} time structure of the time-bin photons. To see this, let us consider the particular case in which only one Schmidt mode is present for a single pulse, {\em{i.e.}}, the JSA factorizes as
\begin{align}
f_p(\omega_s,\omega_i)= \xi_s(\omega_s)\xi_i(\omega_i).
\end{align}
Then, the two-pulse JSA reads
\begin{align}
f_{2p}(\omega_s,\omega_i)&=\xi_s(\omega_s)e^{i\frac{\left(\omega_s-\omega_{s0}\right) T}{2}}\xi_i(\omega_i)e^{i\frac{\left(\omega_i-\omega_{i0}\right) T}{2}}\\ \nonumber
&+\xi_s(\omega_s)e^{-i\frac{\left(\omega_s-\omega_{s0}\right) T}{2}}\xi_i(\omega_i)e^{-i\frac{\left(\omega_i-\omega_{i0}\right) T}{2}}
\end{align}
The output of SPDC in this case can be written as
\begin{align}
\label{eq:eltimebins}
\ket{\phi}=e^{-i gt \left(a_{sE}^{\dagger} a_{iE}^{\dagger}+H.C.\right)}e^{-i gt\left(a_{sL}^{\dagger} a_{iL}^{\dagger}+H.C.\right)}\ket{0}
\end{align}
\noindent with the {\em{early}} and {\em{late}} bins defined as
\begin{align}
\label{eq:ase}
a^{\dagger}_{sE}&=\int d\omega \xi_s(\omega)e^{-i\frac{\left(\omega-\omega_{s0}\right) T} {2}}a^{\dagger}_s(\omega)\\
\label{eq:asl}
a^{\dagger}_{sL}&=\int d\omega \xi_s(\omega)e^{i\frac{\left(\omega-\omega_{s0}\right) T} {2}}a^{\dagger}_s(\omega),
\end{align}
for the signal arm, and as
\begin{align}
\label{eq:aie}
a^{\dagger}_{iE}&=\int d\omega \xi_i(\omega)e^{-i\frac{\left(\omega-\omega_{i0}\right) T} {2}}a^{\dagger}_i(\omega)\\
\label{eq:ail}
a^{\dagger}_{iL}&=\int d\omega \xi_i(\omega)e^{i\frac{\left(\omega-\omega_{i0}\right) T} {2}}a^{\dagger}_i(\omega).
\end{align}
for the idler arm.  In the single photon approximation Eq.~\eqref{eq:eltimebins} yield the state
\begin{align}
\label{eq:eltimebins_approx}
\ket{\phi}=\frac{1}{\sqrt{2}}\left(a^{\dagger}_{sE} a^{\dagger}_{iE}\ket{0}+a^{\dagger}_{sL} a^{\dagger}_{iL}\ket{0}\right)
\end{align}

Note that the {\em{early}} and {\em{late}} bins defined above, Eqs.~\eqref{eq:ase}-\eqref{eq:ail}, are orthogonal only in the limit of large time separation $T$. {Previous theoretical treatments of time-bin experiments have often employed simplified models in which the} {\em{early}} and {\em{late}} labels are treated as good quantum numbers. While this approximation may be adequate in certain regimes, it fails to capture the mixing of {\em{early}} and {\em{late}} bins induced by 1-bit delay measurement interferometers (e.g. an unbalanced Mach-Zehnder interferometer (MZI)), which is central to the evaluation of interference visibility in time-bin entanglement-swapping experiments~\cite{exp_swap_relay2005}. In these simplified models, the two bins effectively reside in different Hilbert spaces. 

In contrast, our approach represents the \emph{early} and \emph{late} bins within a common Hilbert space and expresses them in the same frequency and time bases. Temporal displacements between the bins are then implemented through frequency-dependent phase rotations, providing a unified and self-consistent framework for modeling the action of MZ interferometers and the resulting interference effects. As an example, the average photon number of the temporal modes in arm $2$ immediately downstream of the SPDC source is shown in Fig.\ref{fig:timebins}(b), where the characteristic two-pulse time-bin structure can be observed. Figure~\ref{fig:timebins}(c) shows the corresponding three-pulse time-bin structure in arm $1$ after the MZI.

\begin{figure}[t]
  \centering
\includegraphics[width=\linewidth,trim={0 0 0 0}, clip]{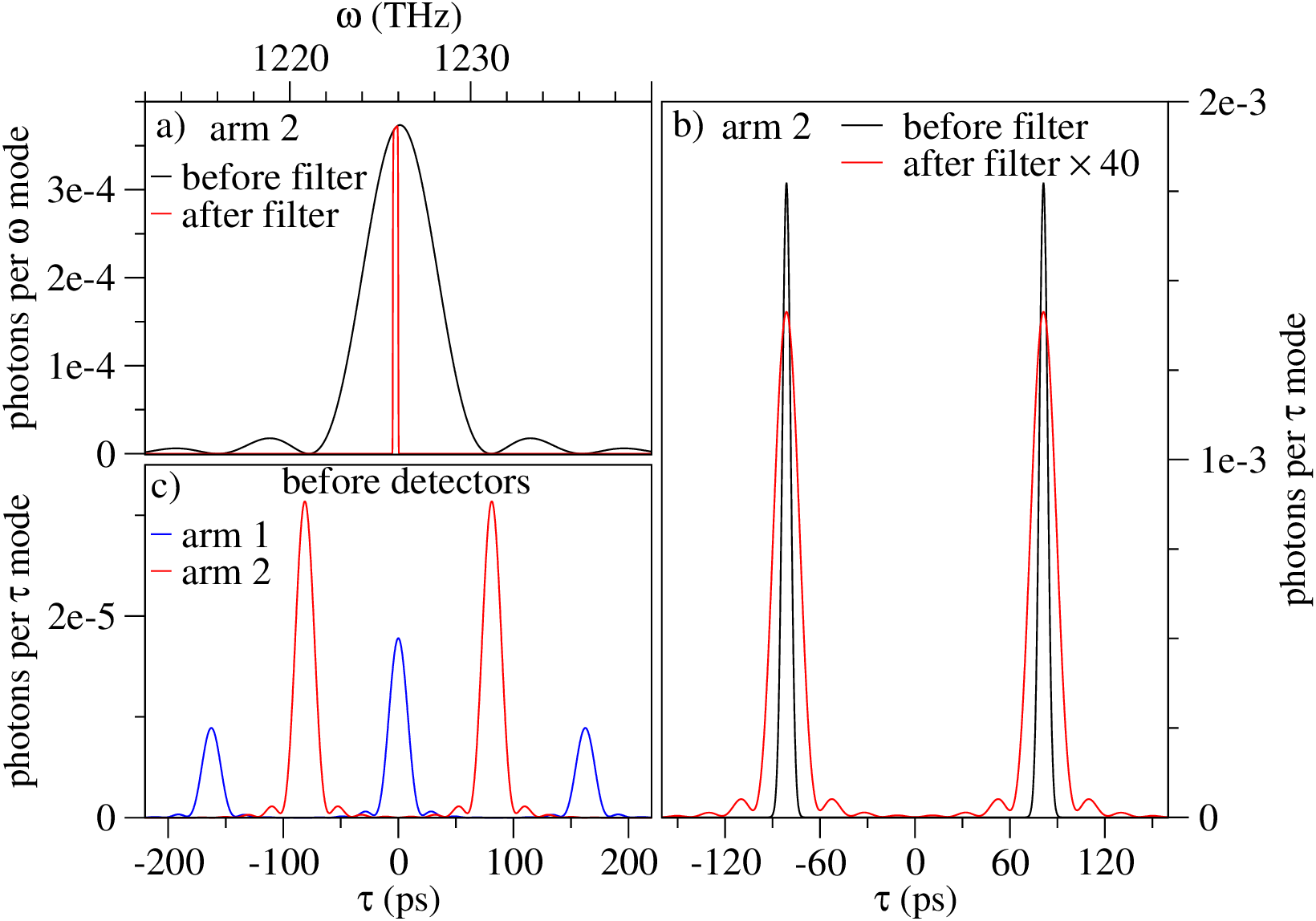}
\setlength{\belowcaptionskip}{-8.5pt} \caption{a) Average number of photons per frequency mode in arm $2$, immediately before and after the filter. b) Average number of photons per temporal mode in arm $2$, immediately before and after the filter. The number of photons after the filter is scaled by a factor of 40 for visibility. c) Average number of photons per temporal mode in arms $1$ and $2$, immediately before the detectors.}
\label{fig:timebins}
\end{figure}

\subsection{Filters}

The filters are modeled as frequency-dependent Gaussian losses~\cite{Branczyk_2010}. This treatment is equivalent to representing the filters as frequency-dependent beam splitters,
\begin{align}
a^{\dagger}_s(\w) &\longrightarrow T_s(\w)a^{\dagger}_s(\w)+R_s(\w)v^{\dagger}_s(\w)\\
a^{\dagger}_i(\w) &\longrightarrow T_i(\w)a^{\dagger}_i(\w)+R_i(\w)v^{\dagger}_i(\w),
\end{align}
\noindent where $v_s(\omega)$ and $v_i(\omega)$ denote vacuum modes. The effect of the filtering is then obtained by tracing over these vacuum modes. The filters are completely characterized by the transmissivity
function $T_{s,i}(\w)$, satisfying 
\begin{align}
\abs{T_{s,i}(\w)}^2+\abs{R_{s,i}(\w)}^2=1.
\end{align}

In the single-photon approximation, the output of the SPDC source is described by the initial state
\begin{align}
\ket{\phi} =\int d\w_1 \int d\w_2 f(\w_1,\w_2)a_s^{\dagger}(\w_1)a_i^{\dagger}(\w_2)\ket{0}.
\end{align}
After filtering, the state becomes a mixed state,
\begin{align} 
\label{eq:rhofilter}
\rho&=\ketbra{\phi^\prime}{\phi^\prime}+C_0\ketbra{0}{0}\\ \nonumber
&+\int d \w_1 \int d \w_2  C_s(\w_1,\w_2) \ a^{\dagger}_s(\w_1)\ketbra{0}{0} a_s(\w_2) \\ \nonumber
&+ \int d \w_1 \int d \w_2  C_i(\w_1,\w_2)  a^{\dagger}_i(\w_1)\ketbra{0}{0} a_i(\w_2) 
\end{align}
\noindent where
\begin{align}
\ket{\phi^\prime} & =\int d\w_1 \int d\w_2 f(\w_1,\w_2)T_s(\w_1)T_i(\w_2)a_s^{\dagger}(\w_1)a_i^{\dagger}(\w_2)\ket{0}\\
C_0&=\int d \w_1 \int d \w_2 \abs{f(\w_1,\w_2)}^2  \abs{R_s (\w_1)}^2 \abs{R_i(\w_2)}^2\\
C_s(\w_1,\w_2)&=T_s(\w_1) T_s^*(\w_2)\int d \w f(\w_1,\w) f^*(\w_2,\w)  \abs{R_i(\w)}^2 \\
C_i(\w_1,\w_2)&=  T_i(\w_1)   T^*_i(\w_2) \int d \w f(\w,\w_1) f^*(\w,\w_2) \abs{R_s(\w)}^2.
\end{align}

Aside from the two-photon component described by the first term in Eq.~\eqref{eq:rhofilter}, the post-filtering density operator contains a vacuum component (the second term), as well as single-photon signal and idler components (the third and fourth terms, respectively).

Using Eq.~\eqref{eq:rhofilter}, the probabilities of detecting a two-photon event, a single signal photon, and a single idler photon are given by
\begin{align}
\Gamma_{2f}&=\norm{\ket{\phi^\prime}}^2=\int d\w_1 \int d\w_2 \abs{f(\w_1,\w_2)}^2\abs{T_s(\w_1)}^2\abs{T_i(\w_2)}^2 \\
\Gamma_{s}&=\Gamma_{2f}+\int d \w C_s(\w,\w)
=\int d \w_1 \int d \w_2
 \abs{f(\w_1,\w_2)}^2 \abs{T_s(\w_1)}^2\\
 \Gamma_{i}&=\Gamma_{2f}+\int d \w C_i(\w,\w)
=\int  d \w_1 \int d \w_2
 \abs{f(\w_1, \w_2)}^2 \abs{T_i(\w_2)}^2.
\end{align}
These probabilities can be used to define a loss-independent heralding efficiency in the single-photon approximation \cite{JSAanalyit_Silberhorn},
\begin{align}
\delta_{s,i}=\frac{\Gamma_{2f}}{\Gamma_{i,s}}.
\end{align}
which characterizes how filtering into the JSA impacts the probability to obtain a coincidence count.

We note that when estimating the mean photon number using experiment, we use this value to correct for the multiphoton impact caused by a lower heralding efficiency. 
This is given by $\mu_{s,i} \approx \delta_{s,i} \times {S_sS_i}/{C}$ \cite{thomas_filt}, where we estimate $\delta_{s,i} \approx 0.5$ accounting for our source's JSAs before and after filtering. In the equation above, $C$ is the two-fold coincidence rate of the photon pair source and $S_s S_i$ is the product of the single photon count rates in each detector.

Before the filters are applied, the output of the SPDC source in our simulations is represented in the Schmidt basis, which accounts for the frequency correlations encoded in the JSA. However, because the action of the filters is frequency dependent, the Schmidt modes are not preserved. Therefore, we transform the state from the Schmidt basis to the frequency basis prior to applying the filters. This transformation is the inverse of Eq.~\eqref{eq:schmidtdecomp} and is implemented using the unitary matrices obtained from the singular value decomposition of the JSA function.
The corresponding update of the Wigner function is then described by a symplectic transformation of the covariance matrix.

The filter is modeled as a loss channel with frequency-dependent transmittance $\abs{T(\omega)}^2$. Consistent with the experimental implementation, rectangular filters are applied to each arm, such that $\abs{T(\omega)}^2=1$ within a narrow frequency range centered at the mean frequency and $\abs{T(\omega)}^2=0$ elsewhere. 
Figure~\ref{fig:timebins}(a) illustrates the average photon number of the frequency modes in arm $2$ immediately before and after filtering.
Figure~\ref{fig:timebins}(b) shows the corresponding photon number of the temporal modes, obtained through a Fourier-transform basis transformation. Note that the rectangular filters not only broaden the time bins but also introduce slowly decaying, long-range oscillatory tails as a consequence of the sharp edges of the rectangular transmission function.

\subsection{Detectors}
\label{supp_detectors}

Defining the Wigner distribution of an operator $O$ as
\begin{align}
\label{eq:WO}
W_{O}(\vec{x},\vec{p})
&=
\left(\frac{1}{2\pi}\right)^N
\int d^N y\,
\opmatrix{\vec{x}-\frac{\vec{y}}{2}}{O}{\vec{x}+\frac{\vec{y}}{2}}
e^{i\vec{y}\cdot\vec{p}},
\end{align}
the Hilbert--Schmidt inner product between two operators can be expressed as an overlap of their Wigner distributions. In particular,
\begin{align}
\label{eq:TRO}
\operatorname{Tr}(\rho O)
&=
(2\pi)^N
\int dx_1\,dp_1\cdots\int dx_N\,dp_N\,
W_{\rho}(\vec{x},\vec{p})W_O(\vec{x},\vec{p}).
\end{align}
This relation provides a convenient way of incorporating on--off photodetection into the Gaussian formalism.

For an ideal single-mode SNSPD, the \emph{off} and \emph{on} outcomes are described by the POVM elements
\begin{align}
\Pi_0 &= \ketbra{0}{0},\\
\Pi_1 &= \mathrm I-\ketbra{0}{0},
\end{align}
respectively. Since the vacuum state has a Gaussian Wigner distribution, probabilities involving $\Pi_0$ reduce to Gaussian overlap integrals and can therefore be evaluated analytically [see also Ref.~\cite{Takeoka_2015}]. For example, for a zero-mean $N$-mode Gaussian state with covariance matrix $V_{2N}$, the probability of projecting all modes onto the vacuum is
\begin{align}
\operatorname{Tr}\!\left(\rho\ketbra{0_N}{0_N}\right)
&=
\frac{2^N}
{\sqrt{det\left(\mathrm I_{2N}+2V_{2N}\right)}}.
\end{align}
Thus, the quantities required to describe on--off detection are directly accessible within a Gaussian representation.

We now generalize this construction to multiple detectors. Let
\begin{align}
\mathcal D &\equiv \{1,\ldots,D\}
\end{align}
denote the set of detectors and
\begin{align}
\mathcal M &\equiv \{1,\ldots,N\}
\end{align}
the full set of optical modes. Detector $r\in\mathcal D$ monitors the subset
\begin{align}
d_r &\equiv \{1_r,\ldots,n_r\}\subseteq\mathcal M.
\end{align}
We assume that different detectors monitor disjoint sets of modes,
\begin{align}
d_r\cap d_{r'}&=\varnothing,
\qquad r\neq r',
\end{align}
so that each mode is assigned to at most one detector. The set of modes not monitored by detector $r$ is denoted by
\begin{align}
\bar d_r &\equiv \mathcal M\setminus d_r.
\end{align}

The \emph{off} outcome of detector $r$ corresponds to all modes in $d_r$ being in the vacuum state. Its POVM element, expressed on the full $N$-mode Hilbert space, is therefore
\begin{align}
\label{eq:detector_off}
\Pi_{r;0}
&=
\left(
\bigotimes_{k\in d_r}\ketbra{0_k}{0_k}
\right)
\otimes \mathrm I_{\bar d_r}.
\end{align}
The \emph{on} outcome is the complementary event in which at least one of the modes monitored by detector $r$ is occupied,
\begin{align}
\label{eq:detector_on}
\Pi_{r;1}
&=
\mathrm I_{\mathcal M}-\Pi_{r;0}.
\end{align}

Consider an $M$-fold coincidence event in which the detectors belonging to the subset
\begin{align}
\mathcal C
&\equiv
\{i_1,\ldots,i_M\}
\subseteq\mathcal D
\end{align}
register clicks, while all remaining detectors register no clicks. We denote the complementary set of detectors by
\begin{align}
\overline{\mathcal C}
&\equiv
\mathcal D\setminus\mathcal C
=
\{j_1,\ldots,j_{D-M}\}.
\end{align}
The probability of observing this detection pattern is
\begin{align}
\label{eq:coincidence_probability}
P(\mathcal C)
&=
\operatorname{Tr}\!\left[
\rho
\left(\prod_{r\in\mathcal C}\Pi_{r;1}\right)
\left(\prod_{r\in\overline{\mathcal C}}\Pi_{r;0}\right)
\right].
\end{align}
Because the detectors act on disjoint mode subsets, the corresponding POVM elements commute.

Direct evaluation of Eq.~\ref{eq:coincidence_probability} involves the \emph{on} operators $\Pi_{r;1}$, whose Wigner representations are not purely Gaussian. However, using $\Pi_{r;1}=\mathrm I_{\mathcal M}-\Pi_{r;0}$ for each clicking detector, the coincidence probability can be expressed entirely in terms of \emph{off} operators. Expanding the product gives
\begin{align}
\label{eq:coincidence_inclusion_exclusion}
P(\mathcal C)
&=
\sum_{\mathcal S\subseteq\mathcal C}
(-1)^{|\mathcal S|}
\operatorname{Tr}\!\left[
\rho
\prod_{r\in\overline{\mathcal C}\cup\mathcal S}
\Pi_{r;0}
\right],
\end{align}
where $\mathcal S$ runs over all subsets of $\mathcal C$ and $|\mathcal S|$ denotes the cardinality of $\mathcal S$.

For an arbitrary subset of detectors $\mathcal A\subseteq\mathcal D$, it is useful to define the joint no-click probability
\begin{align}
\label{eq:P0_definition}
P_0(\mathcal A)
&\equiv
\operatorname{Tr}\!\left[
\rho
\prod_{r\in\mathcal A}\Pi_{r;0}
\right].
\end{align}
Here, the outcomes of detectors outside $\mathcal A$ are unrestricted. In terms of these joint no-click probabilities, the $M$-fold coincidence probability takes the compact form
\begin{align}
\label{eq:coincidence_compact}
P(\mathcal C)
&=
\sum_{\mathcal S\subseteq\mathcal C}
(-1)^{|\mathcal S|}
P_0\!\left(\overline{\mathcal C}\cup\mathcal S\right).
\end{align}
This representation is particularly convenient for Gaussian simulations: each $P_0(\mathcal A)$ involves only projections onto multimode vacuum states, whose Wigner distributions are Gaussian. Consequently, every term contributing to the coincidence probability can be evaluated analytically using Gaussian overlap integrals. 

\subsubsection{Dark counts}

Following Ref.~\cite{Takeoka_2015}, dark counts can be incorporated by modifying the on--off POVM elements as
\begin{align}
\Pi_0\left(\nu\right)
&=
\left(1-\nu\right)\ketbra{0}{0},\\
\Pi_1\left(\nu\right)
&=
\mathrm I-\Pi_0\left(\nu\right),
\end{align}
where $0\leq\nu\leq1$ denotes the probability of a dark count within the detection window. Thus, even when the optical mode is in the vacuum state, the detector produces an \emph{off} outcome only with probability $1-\nu$.

For multiple detectors, allowing each detector $r$ to have an independent dark-count probability $\nu_r$, the joint no-click probability defined in Eq.~\ref{eq:P0_definition} becomes
\begin{align}
\label{eq:P0_definition_nu}
P_0^{\nu}(\mathcal A)
&\equiv
\operatorname{Tr}\!\left[
\rho
\prod_{r\in\mathcal A}
\Pi_{r;0}\left(\nu_r\right)
\right]\nonumber
=
\operatorname{Tr}\!\left[
\rho
\prod_{r\in\mathcal A}
\left(1-\nu_r\right)\Pi_{r;0}
\right]\nonumber
=
\left[
\prod_{r\in\mathcal A}
\left(1-\nu_r\right)
\right]
P_0(\mathcal A).
\end{align}
Hence, dark counts simply rescale the joint no-click probability by the product of the no-dark-count probabilities associated with the detectors in $\mathcal A$.

The  derivation of the $M$-fold coincidence probability remains unchanged. Replacing the ideal joint no-click probabilities by their dark-count-modified counterparts gives
\begin{align}
P^{\nu}(\mathcal C)
&=
\sum_{\mathcal S\subseteq\mathcal C}
(-1)^{|\mathcal S|}
P^{\nu}_0\!\left(
\overline{\mathcal C}\cup\mathcal S
\right).
\end{align}
Thus, dark counts can be incorporated without altering the Gaussian evaluation of the individual terms: they enter only through multiplicative factors associated with the corresponding detector subsets.

\section{Modeling the impact of spontaneous Raman scattering in entanglement swapping experiments}
\label{supp_material_length_sim}

Here we provide an extended discussion of the simulation in the main text which evaluates the relative performance as the fiber length is increased and different quantum-classical wavelength allocations are chosen. 

\paragraph{Raman model:} To model the physics of SpRS in quantum-classical coexistence environments, we extend the approach in Refs.~\cite{thomasx, thomasSPIE_2}. The mean number of SpRS photons at the end of a fiber $l$ arriving within a detection time window $\Delta T_j$ can be modeled by equation~\cite{PhysRevX.2.041010, RamanModel}
\begin{equation}
n_{R_l}=\beta_l\left(\lambda_q, \lambda_{cl}\right) P_{0_l} L_{\mathrm{e f f}_l} \frac{\lambda_q}{h c} \Delta \lambda_j \Delta T_j.
\end{equation}
The linear SpRS coefficient is represented by $\beta_l\left(\lambda_q, \lambda_{cl}\right)$ and is a function of the selections of quantum and classical wavelengths $\lambda_q$ and $\lambda_{cl}$, respectively. 
The classical launch power multiplexed directly into the fiber is $P_{0_l}$. The subscripts $j$ and $l$ index a particular single-photon detector or fiber link, respectively.

For spectrotemporal filtering,
the Raman spectrum can be assumed to be flat across a bandwidth of $<1$~nm. We represent a passband filter in channel $j$ by $\Delta \lambda_j$ (full width at half maximum bandwidth). $L_{e f f_l}$ is the effective interaction length of the Raman interaction, which is dependent on propagation losses for the pump and scattered photons over the long-distance fiber length. This also depends on whether signals co- or counter-propagate in the fiber. These two cases can be modeled using the equations for the effective lengths~\cite{RamanModel}

\begin{equation}
L_{\mathrm{eff_l}}^{\mathrm{co}}=\frac{e^{-\alpha_{cl} L_l}-e^{-\alpha_{c l} L_l}}{\alpha_{cl}-\alpha_{q}},
\end{equation}

\begin{equation}
L_{\mathrm{eff_l}}^{\mathrm{count}}=\frac{e^{\alpha_{cl} L_l}-e^{-\alpha_{q} L_l}}{\alpha_{q}-\alpha_{cl}} e^{-\alpha_{cl} L_l},
\end{equation}

\noindent where $L_l$ is the fiber length and $\alpha_{cl}$ and $\alpha_{q}$ are linear fiber attenuation coefficients representing the classical and quantum signals, respectively.

From our characterization of Raman noise for the 5-km experiments, we estimate $\beta_l\left(1536\,\rm nm, 1547\,\rm nm\right) \approx 3.5 \times 10^{-9} \ \rm{km}^{-1}nm^{-1}$, which agrees with previous studies characterizing C-band/C-band SpRS in standard optical fiber \cite{eraerds_quantum_2010_2, dynes_ultra-high_2016_2, thomasSPIE_2}. When comparing to the O-band/C-band case, we estimate that the SpRS coefficient for $\lambda_q = 1310$~nm and $\lambda_q = 1550$~nm is $\beta_l\left(1310\,\rm nm, 1550\,\rm nm\right) \approx 1.5 \times 10^{-13}\ \rm{km}^{-1}nm^{-1}$. This is roughly four orders of magnitude lower Raman impact as a function of C-band power compared to the allocation used in the experiment. SpRS can be reduced even further using $<1300$~nm channels, which gives a $\sim10^5$ improvement in tolerance to C-band power compared to C-band quantum networking~\cite{Thomas:23, thomasx, thomasSPIE_2, Talcott2026EntanglementFiber}.

A more detailed description of the Raman scattering model and some implications for multiphoton quantum-classical networking can be found in Ref.~\cite{thomasSPIE_2}.

\paragraph{Noise count probability in each detector:}

Relating to the detector model for on-off detectors, the probability to obtain a background count $\nu$ is given by $\nu_j \approx r_j + \mathcal{D}_j$. Here, $r_j$ is the probability per detector window to obtain a noise count due to background noise and $\mathcal{D}_j$ is the dark count probability for single-photon detector $j$. 

For the two swapped arms of the system, $r_j = \eta_{j}n_{R_j}$, where $\eta_j$ is the system efficiency for detecting SpRS photons exiting the fiber. For the BSM detectors, the SpRS noise in each incoming fiber is mixed at a beam splitter, meaning the background noise count has contributions from both. For these detectors, the noise is modeled $r_j = \eta_{j}(n_{R_{1C}} + n_{R_{2C}})$, where the subscripts $1C$ and $2C$ denote the fiber links connecting entangled photons Source 1 and Source 2 to Charlie's BSM node, respectively.

\paragraph{Simulating swapping versus fiber length and wavelength selection:} For the 5-node entanglement swapping experimental configuration, we simulate three key wavelength allocations that would have varying pros and cons in their designs. We simulate the C-band/C-band (C/C) scenario used in this experiment and compare it to the performance of using an O-band/C-band (O/C) or C-band/O-band (C/O) quantum/classical WDM channel assignment. In the C/O case, we keep our SPDC source wavelengths fixed in the C-band (1536.6~nm) but we move the classical signal frequency by $\sim35$-THz to the 1310-nm O-band channel, which shifts the center of the Raman scattering spectrum. This reduces the Raman gain at the quantum wavelengths \cite{Burenkov:23_2, thomas_ofc_2023_2}. On the other hand, the O/C configuration places the quantum signal's WDM channel in the O-band but leaves the classical channel in the C-band. This scheme gives the lowest SpRS contribution by many orders of magnitude due to the suppression of phonon population for anti-Stokes SpRS \cite{Thomas:23, thomasSPIE_2}, but at the cost of higher attenuation ($\sim 0.33$~dB/km) for the quantum photons and therefore lower raw coincidence rates. For all C-band simulations we use a typical fiber attenuation in standard fiber of $\sim 0.2$~dB/km. To be consistent and general, we assume single-mode and indistinguishable SPDC source. We set each source's pump power such that $\mu_{\rm S1=S2} =0.01$. 

The degree of SpRS spectrotemporal filtering is used as in our experimental setup. This keeps our general quantum-classical coexistence designs fixed so that we can evaluate how well these techniques preserve fidelity under different conditions. We also fix the received classical power at the end of each fiber connection to be $P_R = -18$~dBm, slightly above the necessary power for our classical devices to operate error-free, where launch power is increased as a function of fiber length according to added fiber transmission losses to the classical power. 

We find that an O/C system allows high fidelity over distances where a C/C network could not operate due to high noise, showing that O-band quantum systems can reach longer distances despite having higher fiber loss. Notably, the C/C simulation, which is the allocation used in our experiment, can operate over a total distance of $\sim100$~km. Beyond this range, moving the quantum or classical channels to a different wavelength band becomes increasingly advantageous. In principle, an O-band quantum repeater could compensate the sacrifice due to loss compared to the C-band by beating the rate-loss limit for memoryless systems simulated here.

\end{document}